\documentclass[reprint,aps,showpacs,prx,superscriptaddress,floatfix,10pt,nofootinbib]{revtex4-2}
\usepackage[T1]{fontenc}
\usepackage[utf8]{inputenc}
\usepackage{amsmath}
\usepackage{amssymb}
\usepackage{esint}
\usepackage{graphicx}
\usepackage{bbold}
\usepackage{appendix}
\usepackage{bibunits}
\usepackage{color,xcolor}
\usepackage{physics} 
\usepackage{verbatim}
\usepackage{bm}

\defaultbibliographystyle{myapsrev4-2}

\usepackage{braket, orcidlink, amssymb, amsmath, dsfont, adjustbox, multirow, lineno, nccmath, scalerel, mathtools, booktabs}

\PassOptionsToPackage{breaklinks=true,colorlinks=true}{hyperref}
\PassOptionsToPackage{hyphens}{url}

\usepackage{ragged2e}
\usepackage{caption}
\usepackage{algorithm}
\usepackage{algpseudocode}

\usepackage[normalem]{ulem}

\usepackage{dcolumn}

\usepackage{tikz}
\usetikzlibrary{quantikz2}

\hypersetup{
    unicode=true,
    pdftoolbar=true,
    pdfmenubar=true,
    pdffitwindow=false,
    pdfstartview={fitH},
    pdftitle={},
    pdfauthor={A.~Chakraborty, B.~Sambasivam, K.~Shirali, M.~Ramôa, H. Nelson, S.E.~Economou, E.~Barnes},
    pdfsubject={},
    pdfcreator={A.~Chakraborty, B.~Sambasivam, K.~Shirali, M.~Ramôa, H. Nelson, S.E.~Economou, E.~Barnes},
    pdfproducer={},
    pdfkeywords={},
    pdfnewwindow=true,
    colorlinks=true,
    breaklinks=true,
    linkcolor=blue,
    citecolor=magenta,
    filecolor=magenta,
    urlcolor=blue
}

\makeatletter
\def\@fnsymbol#1{\ensuremath{\ifcase#1\or \dagger\or \ddagger\or
   \mathsection\or \mathparagraph\or \|\or **\or \dagger\dagger
   \or \ddagger\ddagger \else\@ctrerr\fi}}
    \makeatother
\makeatother

\begin{document}
\title{An efficient algorithm for approximate shadow Hamiltonian simulation}

\author{Abhijit Chakraborty$^*$\orcidlink{0000-0001-7487-9925}}
\email{abhijit.iiserk@gmail.com}
\affiliation{Department of Physics, Virginia Tech, Blacksburg, VA 24061, USA}
\affiliation{Virginia Tech Center for Quantum Information Science and Engineering, Blacksburg, VA 24060, USA}

\author{Bharath Sambasivam$^*$\orcidlink{0000-0002-5765-9469}}
\email{sbharath@vt.edu}
\affiliation{Department of Physics, Virginia Tech, Blacksburg, VA 24061, USA}
\affiliation{Virginia Tech Center for Quantum Information Science and Engineering, Blacksburg, VA 24060, USA}

\author{Karunya Shirali\orcidlink{0000-0002-2006-2343}}
\affiliation{Department of Physics, Virginia Tech, Blacksburg, VA 24061, USA}
\affiliation{Virginia Tech Center for Quantum Information Science and Engineering, Blacksburg, VA 24060, USA}

\author{Hunter Nelson\orcidlink{0009-0002-8560-5385}}
\affiliation{Department of Physics, Virginia Tech, Blacksburg, VA 24061, USA}
\affiliation{Virginia Tech Center for Quantum Information Science and Engineering, Blacksburg, VA 24060, USA}

\author{Mafalda Ram\^{o}a\orcidlink{0000-0003-0218-7801}}
\affiliation{Department of 
Physics, Virginia Tech, Blacksburg, VA 24061, USA}
\affiliation{Virginia Tech Center for Quantum Information Science and Engineering, Blacksburg, VA 24060, USA}

\author{Sophia E. Economou\orcidlink{0000-0002-1939-5589}}
\affiliation{Department of Physics, Virginia Tech, Blacksburg, VA 24061, USA}
\affiliation{Virginia Tech Center for Quantum Information Science and Engineering, Blacksburg, VA 24060, USA}

\author{Edwin Barnes\orcidlink{0000-0002-0982-9339}}
\email{efbarnes@vt.edu}
\affiliation{Department of Physics, Virginia Tech, Blacksburg, VA 24061, USA}
\affiliation{Virginia Tech Center for Quantum Information Science and Engineering, Blacksburg, VA 24060, USA}

\date{\today}

\def\thefootnote{*}\footnotetext{These authors contributed equally to this work.}

\begin{bibunit}

\begin{abstract}
We propose an efficient algorithm based on shadow Hamiltonian simulation
to approximately simulate the real-time dynamics of observables under time-independent Hamiltonians. Shadow Hamiltonian simulation works at the level of the operator algebra generated by the observables through commutators with the Hamiltonian. Exactly encoding the quantum state in this picture is generally inefficient for interacting systems due to the exponential growth of the operator algebra. Our algorithm overcomes this bottleneck by systematically identifying the elements of the algebra most relevant to the target observables. This targeted approach is a controlled approximation that yields a highly efficient quantum state encoding that substantially reduces the size of the qubit register required to perform the time evolution using the shadow Hamiltonian. We propose two main pruning schemes, one based on a predefined operator basis and another on a constructed Krylov basis. We also present a hybrid scheme that builds a Krylov basis within a pruned algebra in the predefined basis. We benchmark our algorithm using lattice spin systems in one and two dimensions, for both one- and higher-point correlators as observables.

\end{abstract}


\maketitle

\section*{Introduction}
The study of the real-time dynamics of many-body Hamiltonians is a problem of central importance in a wide range of fields, including high-energy and nuclear physics~\cite{JordanQuantum2018,BanulsEPJD2020,DavoudiQuantum2024,Bennewitz2024}, materials science, condensed matter physics~\cite{KarraschPRB2012,GustafsonPRD2019,BulchandaniPNAS2023,EcksteinNPJQI2024,LeePRL2026}, quantum chaos and thermalization~\cite{JunPRX2017,YangPRXQuantum2023,HahnPRX2024,AnandScientificReports2024,AsaduzzamanPRD2024,PerrinNatrureComm2025}, quantum chemistry and molecular dynamics~\cite{KassalNASProceedings2008,ChanScience2023,KaleJPhysChemLett2024}. The rapid growth of entanglement in the quantum state creates a bottleneck for classical computing techniques for Hamiltonian simulation, such as tensor networks~\cite{OrusNatureReviews2019}, Pauli propagation~\cite{Rudolph2025}, and lightcone-based methods~\cite{PerezPRB2022,KimNature023}. Developing efficient quantum algorithms to overcome this challenge may lead to a potential quantum advantage for problems of interest. Existing quantum algorithms that simulate the evolution of the quantum state comprise approaches based on product formulae~\cite{SuzukiJMP1991,BerryCMP2007,PoulinPRL2011,WiebeJPhysA2011,ChildsQuantum2019,ChildsPhysRevX2021}, randomized channels~\cite{CampbellPRL2019,NakajiPRXQ2024}, linear combination of unitaries~\cite{Childs2012,ChakrabortyQuantum2024}, quantum signal processing~\cite{LowPRL2017,AlexisSpringerScience2024}, and variational approaches~\cite{YuanQuantum2019,CirstoiuNPJQI2020,YaoPRXQ2021,KokcuPRL2022,Alsheikh2025,Sambasivam2026}. All these algorithms have their own advantages and disadvantages, either in terms of asymptotic scaling of resources in time and error, or in terms of their near-term feasibility. A common feature of these quantum algorithms is the usage of a direct encoding of an $N$-qubit state onto a physical register of $N$ qubits.

In many cases, the ultimate goal of Hamiltonian simulation is to track the evolution of a small set of relevant observables. For this purpose, encoding the full quantum state may not be necessary. With this motivation, Ref.~\cite{SommaNatComm2025} introduced the shadow Hamiltonian simulation algorithm, which works at the level of the algebra generated by the set of observables of interest via commutation with the Hamiltonian. The normalized vector of expectation values of the state on this generated algebra---the shadow state---is encoded onto qubits, and a corresponding shadow Hamiltonian is constructed to simulate its evolution, giving direct access to the evolved expectation value(s). This encoding blue, which can be viewed as mapping a sparse spin vector onto qubits, was shown to be more efficient for non-interacting systems in Ref.~\cite{SommaNatComm2025}, owing to the polynomial growth of their algebras for some classes of observables. However, for interacting systems, this is not the case, with up to $2N$ qubits required to encode the shadow state \emph{exactly} for an $N$-qubit system. Therefore, the primary gain of a more efficient qubit encoding is lost for interesting problems in physics and chemistry, leaving open the question of whether shadow Hamiltonian simulation is practically relevant as a quantum algorithm.

In this work, we present a framework for the approximate quantum simulation of real-time dynamics of observables evolving under a many-body Hamiltonian. Our approach builds upon the principles of shadow Hamiltonian simulation, extending its utility to the practically relevant case of interacting Hamiltonians, with potentially exponentially scaling algebras. We propose an algorithm that classically constructs a truncated shadow Hamiltonian by systematically identifying the operator subspaces most pertinent to a specified set of observables. By initializing the construction from these target observables, the algorithm generates an orthonormal operator basis whose cardinality is governed by the prescribed simulation fidelity and temporal evolution; this yields an approximate shadow state with a dimensionality that can be significantly lower than that of the full Hilbert space. For various lattice models, this methodology facilitates the evolution of expectation values within a drastically reduced effective Hilbert space while preserving controllable accuracy, thereby enabling a more efficient qubit encoding of the shadow state and its associated Hamiltonian. For a 1D mixed-field Ising model with a moderate transverse field, we demonstrate the capability to track the magnetization dynamics of a 100-qubit physical system using only 10 qubits. Numerical evidence suggests that these efficiency gains persist in 2D systems and across diverse lattice models.

To create the truncated orthonormal operator basis starting from the desired observable(s) of interest and subsequently the truncated shadow Hamiltonian, we propose \textit{operator pruning}, for which we present three tractable methods, with their relative merits evaluated: (i) a graph-based approach that constructs the reduced space in a predefined orthonormal operator basis, (ii) an operator Lanczos approach that adaptively builds a truncated Krylov basis, and (iii) a hybrid method that combines favorable aspects of both methods. The truncation method in all cases depends on the Hamiltonian, the target observable, and the desired simulation accuracy.

While the primary intention of this work is to perform Hamiltonian simulation on quantum computers, our classical pre-processing algorithms can also work as a method for classical simulation of dynamics, depending on the size and complexity of the resulting shadow Hamiltonian. There are broadly three regimes in the dimension of the shadow Hamiltonian---first, where the dimension is low enough for exact diagonalization to be tractable; second, where the dimension is too large for exact diagonalization, but the structure of the shadow Hamiltonian allows for the efficient application of classical simulation methods such as tensor networks and Pauli propagation; third, the dimension is too large for exact diagonalization, and the structure of the shadow Hamiltonian is not amenable to classical simulation techniques, leading to the need for quantum algorithms.

We demonstrate the performance of the framework for spin-$1/2$ lattice models in one and two spatial dimensions. Beyond local observables, we show that it accurately captures the dynamics of unequal-time higher-order correlation functions, including current autocorrelation functions relevant to spin transport and out-of-time-ordered correlators (OTOCs) that probe quantum chaos.


\section*{Results}

\begin{figure*}[ht!]
    \centering
    \includegraphics[width=\linewidth]{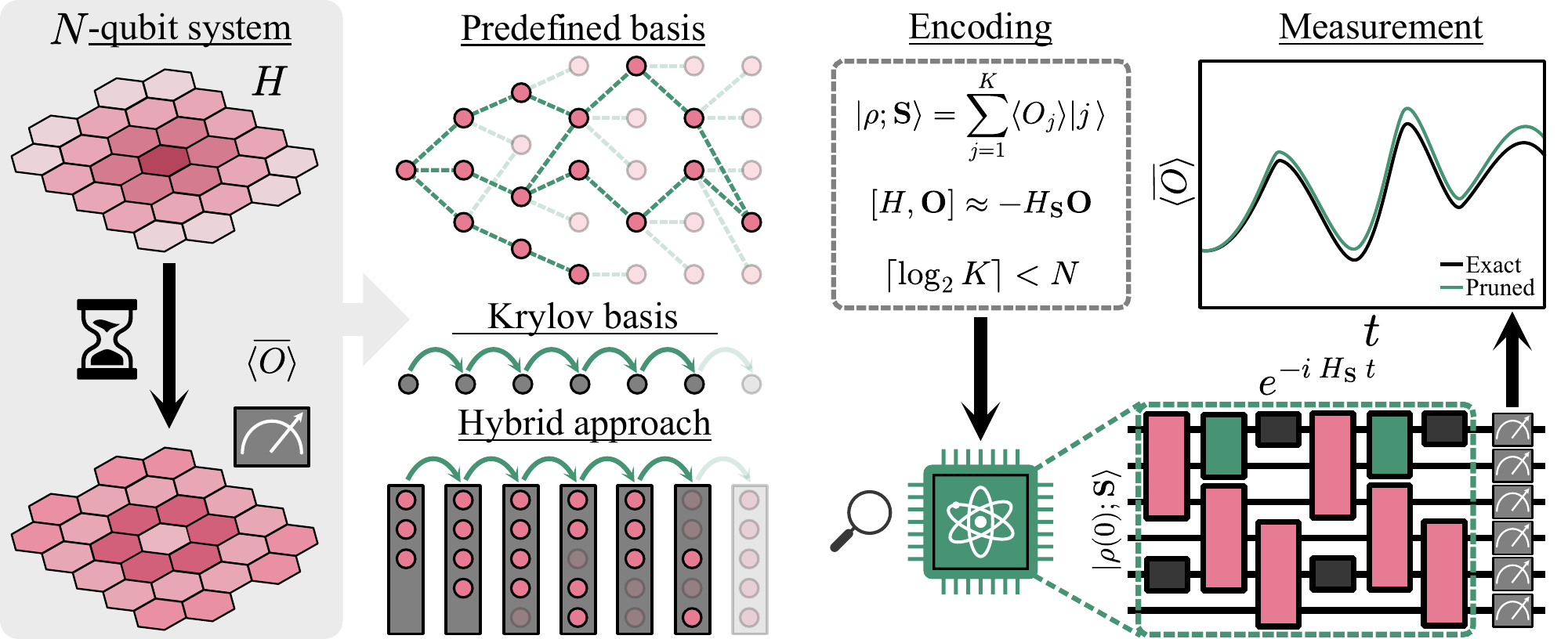}
    \caption{Schematic of the pruned shadow Hamiltonian simulation algorithm to approximately simulate the evolution of the observable $\bar{O}$ under an $N$-qubit Hamiltonian, $H$. Three schemes (predefined basis, Krylov basis, and hybrid) can be used to build a shadow operator basis $\mathbf{S}$ containing important operators to $\bar{O}$ under commutation with $H$. The shadow state encodes the expectation values of these observables on the physical state and can provide a more efficient mapping onto $\lceil\log_2K\rceil<N$ qubits. The shadow state is evolved by the shadow Hamiltonian $H_{\mathbf{S}}$, and measured in the computational basis to obtain the approximate real-time evolution of $\bar{O}$.}
    \label{fig:Schematic}
\end{figure*}


\subsection*{Approximate shadow Hamiltonian simulation}\label{sec:approx-shadow-sim}

Shadow Hamiltonian simulation, introduced in Ref.~\cite{SommaNatComm2025}, is a quantum algorithm that encodes a quantum state as a normalized vector of its expectation values of operators in an algebra to directly track their evolution. This ``shadow state'' is defined with respect to a set $\mathbf{S}\equiv\{O_j\}_{j=1}^K$ which contains the observable(s) of interest:
\begin{equation}\label{eq:ShadowStateDef}
    \ket{\rho;\mathbf{S}}=\frac{1}{\sqrt{A}}\sum_{j=1}^K\langle O_j\rangle\ket{j},
\end{equation}
where all expectation values are on the quantum state $\rho$, $A=\sum_{j}\lvert\langle O_j\rangle\rvert^2$ normalizes $\ket{\rho;\mathbf{S}}$, and $\ket{k}$ is the computational basis state defined by the binary representation of the integer $k-1$. For the task of time-evolving the initial quantum state $\rho(0)$ by the time-independent Hamiltonian $H$, the encoding defined in Eq.~\eqref{eq:ShadowStateDef} into $\ket{\rho(0);\mathbf{S}}$ is used. For the unique determination of the shadow state at a later time, $\ket{\rho(t);\mathbf{S}}$, the set $\mathbf{S}$ must satisfy an invariance property under commutation with $H$: 
\begin{equation}\label{eq:InvarianceProp}
    \left[H,O_j\right]=-\sum_{j'=1}^Kh_{jj'}O_{j'},\quad\forall j=1,\cdots K,
\end{equation}
where $h_{jj'}\in \mathbb{C}$. The shadow state $\ket{\rho(t);\mathbf{S}}$ satisfies the Schr\"{o}dinger equation defined by the \emph{shadow Hamiltonian} $H_{\mathbf{S}}$, defined as the $K\times K$ matrix of coefficients $h_{jj'}$. $H_{\mathbf{S}}$ is Hermitian if the operators in $\mathbf{S}$ are orthogonal under the Hilbert-Schmidt distance:
\begin{equation}
    \Tr\left(O_j^{\dagger}O_{j'}\right)=\lambda\,\delta_{jj'},\quad\forall j,j'=1,\cdots,K,
\end{equation}
where $\lambda>0$. Using this approach, one can study the evolution of expectation values of observables directly by measuring the shadow state in the computational basis.

Here, we introduce our algorithm for performing approximate shadow Hamiltonian simulation. The primary roadblock in the practical use of shadow Hamiltonian simulation is the growth of $\dim(\mathbf{S})$ such that it \emph{exactly} satisfies the invariance property defined in Eq.~(\ref{eq:InvarianceProp}). As shown in Ref.~\cite{SommaNatComm2025}, for \emph{purely non-interacting systems} such as free-fermionic and free-qubit models, this growth is $\mathcal{O}(\text{poly}(N))$, where $N$ is the number of degrees of freedom. This leads to efficient encoding of the shadow state (and Hamiltonian) onto $\mathcal{O}(\log_2\text{poly}(N))$ qubits. However, for interacting systems, $\dim(\mathbf{S})$ generally grows exponentially with $N$ ~\cite{Aguilar2024,Wiersema2024}. This exponential scaling typically drives the cost of shadow simulation up to $2N$ qubits, eliminating any advantage over regular Hamiltonian simulation.

For the task of tracking the time-evolution of a Hermitian observable $\bar{O}$ under $H$, not all operators in its closure $\mathbf{S}$ are equally important. We will refer to $\bar{O}$ as the root observable. Here, we introduce a systematic numerical pruning algorithm with three related realizations that builds a smaller set $\mathbf{S}_{\bar{O}}\subset\mathbf{S}$ of the most important operators. In the case where the evolution of a set of observables $\bar{\mathbf{O}}$ is of interest, the total pruned set is simply the union $\mathbf{S}_{\bar{\mathbf{O}}}\equiv\bigcup_{\bar{O}\in\bar{\mathbf{O}}}\mathbf{S}_{\bar{O}}$, or alternatively parallel simulations for each $\bar{O}\in\bar{\mathbf{O}}$ can be performed with shadow Hamiltonians constructed from the corresponding sets $\mathbf{S}_{\bar{O}}$.

Consider the Heisenberg evolution of $\bar{O}$ written as a series of nested commutators:
\begin{equation}\label{eq:HeisenbergEvol}
    \bar{O}(t)=\sum_{j=0}^{\infty}\frac{(i\,t)^j}{j!}\mathcal{A}_H^{j}(\bar{O}),
\end{equation}
where the superoperator $\mathcal{A}_H(\,\cdot\,):=\left[H,\,\cdot\,\right]\,$. For a complete set of orthonormal observables $\mathbf{S}\ni \bar{O}$ that satisfies Eq.~\eqref{eq:InvarianceProp},
\begin{equation}
    \mathcal{A}_H^j(\bar{O})=\sum_{O\in\mathbf{S}}\alpha^{(j)}_O O,\quad\forall j,
\end{equation}
where $i\,\alpha^{(j)}_O\in\mathbb{R}\,\forall j>0$, and $\alpha^{(0)}_O=1$. Our pruning algorithm introduces an approximation:
\begin{equation}  
\mathcal{A}_H^j(\bar{O})\approx\sum_{O\in\mathbf{S}_{\bar{O}}}\alpha^{(j)}_O O,\quad\forall j,
\end{equation}
where the error stems from $O\in \mathbf{S}\setminus\mathbf{S}_{\bar{O}}$. This also results in an approximate fulfillment of the invariance property in Eq.~\eqref{eq:InvarianceProp}, ensuring that the Schrödinger equation governed by the Hamiltonian
$H_{\mathbf{S}}$ is approximately satisfied. The shadow Hamiltonian corresponding to this approximately complete set will still be Hermitian provided the elements of $\mathbf{S}_{\bar{O}}$ are orthonormal. The Hermitian property of $H_{\mathbf{S}}$ is important for simulating the system on a quantum computer without the need for block encoding.

Our goal is to identify a set of important orthonormal operators that span the algebra generated by the nested commutators of the root observable with the Hamiltonian, $\mathcal{A}^{j}_H(\bar{O})$. Since the efficiency of the classical preprocessing routine depends on the classical cost of constructing and storing these operators, we develop three different approaches for building a truncated basis, which focus on reducing both the classical preprocessing cost and the required qubit register size for shadow Hamiltonian simulation. The first selects the most relevant operators from a predefined orthonormal basis (the Pauli basis in our examples) using a graph-based procedure, where nodes represent operators and edges quantify their relevance to the root observable. The second iteratively constructs an orthonormal Krylov basis using an operator Lanczos algorithm. Finally, we combine these strategies into a hybrid approach that inherits the advantages of both. An overview of the methodology is shown in Fig.~\ref{fig:Schematic}.


\subsubsection{Predefined basis}\label{sec:predefined-basis}
Consider an orthonormal operator basis $\mathbf{B}$ that spans the relevant algebra corresponding to the target Hamiltonian. For example, for an $N$-qubit system, one could consider the basis of all traceless Pauli strings spanning $\mathfrak{su}(2^{N})$. For a fermionic system, one could choose a Majorana basis for the Clifford algebra. Without loss of generality, we will assume that the coefficients of the Hamiltonian expanded in $\mathbf{B}$ are defined in units of an energy scale such that they are upper bounded by 1. 

The process of identifying important operators in $\mathbf{B}$ can be seen as building a weighted layered graph $\left(V,E,w\right)$, where the vertices $V\in\mathbf{B}$ are connected by edges in $E$ with weights given by $w$. The components of $\bar{O}$ in the basis $\mathbf{B}$ are the vertices in the root layer $\{V_{0}\}$. The vertices of subsequent layers of this graph are the basis elements that appear in the action of $\mathcal{A}_H$ on the vertices of the preceding layer:
\begin{equation}
    \{V_{i+1}\} = \{v_{i+1}\in\mathcal{A}_H(v_i)\,\vert\,\forall v_i\in\{V_{i}\}\}.
\end{equation}
A pair of vertices is connected by an edge if an action of $\mathcal{A}_H$ can take one to the other. The weight of an edge $(v_i,v_i+1)\in E$ is the coefficient in the commutator of the basis element to which the resultant vertex corresponds: 
\begin{equation}
    w(v_i,v_{i+1})=\Big\vert\Tr\left(\mathcal{A}_H(v_i)\,v^{\dagger}_{i+1}\right)\Big\vert,
\end{equation}
where $v_{i}\in \{V_{i}\},\,v_{i+1}\in \{V_{i+1}\}$. If a particular basis operator appears again during the graph generation procedure with a lower associated weight than before, we do not add a vertex to the graph. If, on the other hand, it reappears with a larger weight, the new vertex, along with all edges and vertices that get generated from it, replaces the previous corresponding ones with a lower weight. We define the accumulated weight of a particular vertex in the graph as the maximum product of weights along paths connecting the vertex to the root layer
\begin{equation}\label{eq:AccumWts}
    \Omega(v_k)=\max_{\{V_0\}\to v_k}\prod_{i=0}^{k-2} w(v_i,v_{i+1}).
\end{equation}
If $\Omega(v_k)<\epsilon$, where $\epsilon$ is a chosen threshold value, $v_k$ and all edges connected to it are not added to the graph, and the corresponding operators are not part of the approximate shadow Hamiltonian simulation. We provide the pseudocode for this algorithm in Methods~\ref{app:pseudocodes}. The scalability of this method is limited by the classical memory required to store the orthonormal basis elements (nodes of the pruned graph), which depends on the system size, Hamiltonian parameters, the target observable, and the chosen truncation threshold. The number of basis vectors in the set $\mathbf{S}_{\bar{O}}$ also determines the cost of initializing the shadow state according to Eq.~(\ref{eq:ShadowStateDef}).

A natural application of this pruning approach is for perturbative Hamiltonians of the form $H=H_0+\xi\,H_1$, where $\xi$ is a small parameter compared to $\| H_0\|$. The choice of cutoff $\epsilon=\xi^M$, where $M\in\mathbb{Z}_+$, amounts to a $M^{\text{th}}$-order truncation of the Dyson series in the interaction picture corresponding to the Heisenberg evolution in Eq.~\eqref{eq:HeisenbergEvol}. We derive bounds on the error and asymptotic scaling of $M$ in Methods~\ref{app:perturbative-pruning}, along with numerical tests of their tightness in Sec.~\ref{sec:Results}. The worst-case scaling of the complexity of the classical preprocessing is a polynomial with its degree determined by the number of layers $l$ in the graph that needs to be kept for a given threshold:
\begin{align}
    \mathcal{O}\left(\sum_{j=0}^M\binom{l}{j}N_0^{l-j}N_1^j\right),
\end{align}
where $N_{0,1}$ are the number of terms in $H_{0,1}$ when expanded in the basis $B$. We show more details in Methods~\ref{app:ClassComplex}.


\subsubsection{Krylov basis}\label{sec:constructed-basis}
An alternative approach is to construct a Krylov operator basis iteratively informed by the orbit of orthogonal operators stemming from the root observable \cite{NandyPhyRep2025}. Here, we introduce a truncated operator Lanczos algorithm that constructs the Krylov basis, $\mathbf{S}_{\bar{O}}$. For simplicity, we will again assume that there is a single root observable, $\bar{O}$ and that it is normalized, $\|\bar{O}\|=1$. In the first iteration, a new operator is generated using $\tilde{O}_1=-i\,\mathcal{A}_H(\bar{O})$, which is orthogonal to $\bar{O}$ by construction. Then it is normalized and added to the shadow set: $O_1\equiv \tilde{O}_1/\|\tilde{O}_1\|\in\mathbf{S}_{\bar{O}}$. On iteration $k$ of the algorithm, the superoperator is applied to the previous orthonormal operator $O_{k-1}\in\mathbf{S}_{\bar{O}}$:
\begin{equation}
    \tilde{O}_k\equiv-i\,\frac{\mathcal{A}_H(O_{k-1})}{\|\mathcal{A}_H(O_{k-1})\|},
\end{equation}
which can be written as:
\begin{equation}
    \tilde{O}_k=\sum_{O\in\mathbf{S}_{\bar{O}}}c_OO + O^{\perp}_{k},
\end{equation}
where $c_O$ are the coefficients of the linearly dependent part, and $O^{\perp}_{k}$ is a component of $\tilde{O}_k$ that is orthogonal to the current $\mathbf{S}_{{\bar{O}}}$. This component is normalized and added to the shadow set $O_k\equiv O^{\perp}_k/\|O^{\perp}_{k}\|\in\mathbf{S}_{\bar{O}}$, provided the product of its norm with that of all previous orthogonal components is above a preset threshold:
\begin{equation}
    w=\prod_{j=1}^k \|O^{\perp}_{j}\| \geq \epsilon. \label{eq:pruning-condition-Krylov}
\end{equation}
The algorithm terminates if this condition is not met. The multiplicative truncation criterion introduced here differs from the conventional criterion, in which the Krylov iteration terminates once the norm of the orthogonalized vector satisfies $\|O_k^\perp\| < \epsilon$ \cite{NandyPhyRep2025}. By accumulating the relative weight along the entire Krylov trajectory, the multiplicative criterion more effectively identifies operators that remain important with respect to the root operator.

At the conclusion of this classical subroutine, the set of operators $\{O_k\}$ forms an orthonormal Krylov basis up to this truncation threshold $\epsilon$. We provide the pseudocode for this algorithm in Methods~\ref{app:pseudocodes}, and derive a bound on error and an asymptotic scaling of $\epsilon$ given a target precision $\delta$ at time $t$ in Methods~\ref{app:krylov-pruning}.

Although the Krylov pruning algorithm is formally independent of the basis chosen to represent the Krylov vectors, it is convenient to work in a basis in which the target observable and Hamiltonian are sparse for practical computation of the commutators. Despite the initial sparsity, the Krylov vectors in this basis can become progressively dense as the Krylov subspace is expanded. For the systems considered in this work, for example, it is natural to consider the Pauli basis. In this case, an individual Krylov vector can contain contributions from all $4^N$ Pauli strings, resulting in an exponential classical memory requirement to store the vectors. Moreover, constructing the shadow state defined in Eq.~(\ref{eq:ShadowStateDef}) requires evaluating the expectation values of all basis operators present in the Krylov vectors. Consequently, if the full set of $4^N$ Pauli strings contributes, the cost of defining the initial shadow state also scales exponentially with the size of the system. These potentially exponential resource requirements significantly restrict the applicability of the Krylov method beyond moderate system sizes in all the cases considered here. That being said, there could be ways to get around this poor scaling for some problems and observables that admit efficient computation of commutators in a more viable basis. In this work, to reduce the memory overhead, we introduce a hybrid approach that combines graph-based and Krylov-based pruning techniques.


\subsubsection{Hybrid approach}
In this hybrid method, instead of building the Krylov basis on the complete algebra generated by the nested commutators with the Hamiltonian, whose dimension scales exponentially for interacting systems, the basis is built on a subspace defined by the set $\mathbf{S}_{\bar{O}}$ determined by the predefined pruning method for a given threshold. This ensures that the resulting Krylov vectors cannot have more than $|\mathbf{S}_{\bar{O}}|$ elements of the predefined basis, which scales polynomially in system size for a given threshold. Building the pruned Krylov basis as defined in the previous section in this pruned algebra also provides an advantage over the predefined basis method, as we numerically find for the cases we tried in Sec.~\ref{sec:Results} that the number of Krylov basis vectors is much smaller compared to the number of elements in the set $\mathbf{S}_{\bar{O}}$ for the same simulation accuracy. This further reduces the required register size for encoding the shadow Hamiltonian while keeping the cost for shadow state initialization upper bounded by the number of elements appearing in $\mathbf{S}_{\bar{O}}$.

The reduction in required register size and classical cost in defining the Krylov vectors in this hybrid method can be attributed to the two-step pruning method corresponding to two different threshold values $\epsilon \equiv (\epsilon_1, \epsilon_2)$. $\epsilon_1$ is used to obtain the pruned set $\mathbf{S}_{\bar{O}}$ using the predefined basis method. Once the pruned algebra is defined, at each step of evaluating the commutator, the resulting operator $\mathcal{A}_H(O_{k-1})$ is projected on to the set $\mathbf{S}_{\bar{O}}$ and a Gram-Schmidt orthonormalization procedure is performed on this new vector to find the orthogonal component that resides in the same space defined by the span of the basis vectors in $\mathbf{S}_{\bar{O}}$. The construction of this basis continues until Eq.~(\ref{eq:pruning-condition-Krylov}) is no longer satisfied. For the multiplicative norm tracker in Eq.~(\ref{eq:pruning-condition-Krylov}) the threshold is set to $\epsilon_2$, which can generally be different from $\epsilon_1$. A pseudocode for this algorithm is presented in Methods~\ref{app:pseudocodes}. In this article, we only provide results for the Krylov basis for a couple of examples we consider in Sec.~\ref{sec:Results}; due to poor scaling of the Krylov-basis method while choosing the Pauli basis to represent the Krylov vectors in and the overall efficiency of the hybrid method, all further examples include results using the predefined basis pruning and hybrid pruning methods, but not the purely Krylov-basis method.


\subsection*{Numerical results}\label{sec:Results}

\begin{figure*}
    \centering\includegraphics[width=1.0\linewidth]{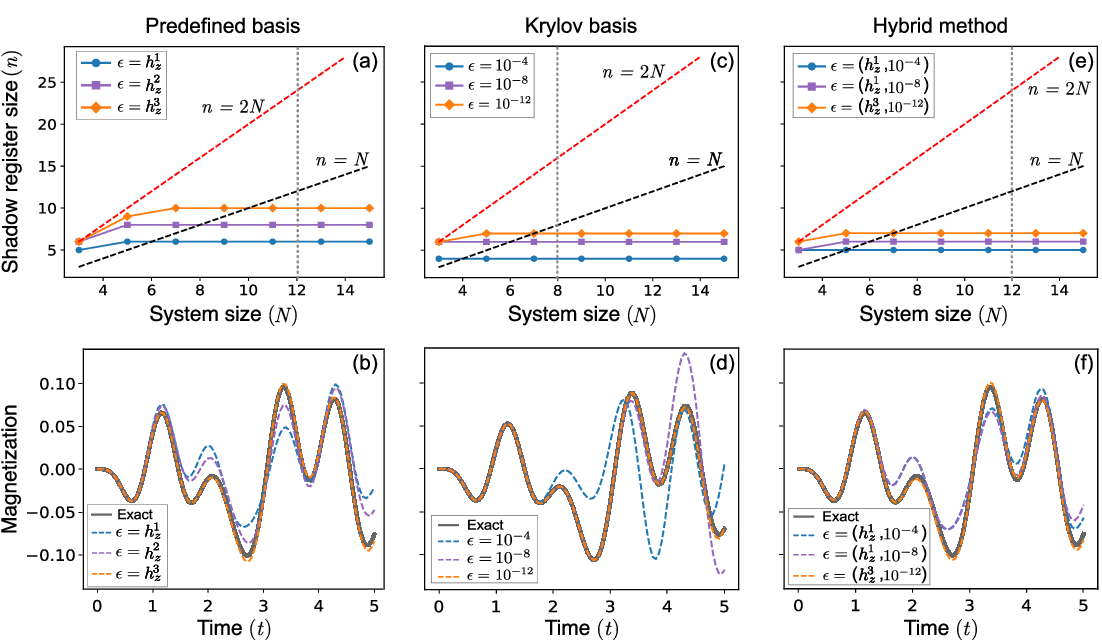}
    \caption{Pruned shadow Hamiltonian simulation of the 1D MFIM with $h_z=0.3$. (a,b) Results for the predefined Pauli-basis pruning scheme. Panel (a) shows the required shadow-register size as a function of system size for different pruning thresholds $\epsilon$, using the central-site $Z$ operator as the target observable. The saturation of the register size indicates that, for fixed $\epsilon$, the magnetization dynamics of arbitrarily large systems can be simulated with a constant number of register qubits. The red dashed line indicates the scaling of the number of qubits needed to perform exact shadow Hamiltonian simulation. The gray dotted line marks the system size used in panel (b), which shows the magnetization dynamics of an $N=12$ chain initialized in $\ket{0}^{\otimes N/2}\otimes\ket{1}^{\otimes N/2}$. (c,d) Corresponding results for the Krylov-basis pruning scheme. For $\epsilon=10^{-12}$, a 7-qubit register is sufficient to simulate a physical system with $N=8$ system (gray dotted line), although the cost of representing each Krylov vector grows exponentially with system size. (e,f) Results for the hybrid scheme, where $\epsilon$ denotes the pair of thresholds used in the Pauli-basis and Krylov-basis pruning stages. By applying the Krylov construction only within the pruned set $\mathbf{S}_{\bar{O}}$, the method remains scalable beyond the system sizes accessible to the pure Krylov approach. For comparable accuracy, the hybrid method reproduces the magnetization dynamics of $N=12$ spins using only $7$ register qubits, compared to $10$ qubits required by the predefined-basis scheme. A complementary absolute error in magnetization vs time plot is shown in Supplementary Information~\ref{app:error-magnetization}.}
    \label{fig:MFIM_main_text}
\end{figure*} 


As a demonstration of the effectiveness of our algorithm, we apply it to various lattice models that exhibit exponential growth for the complete shadow set $\mathbf{S}$ with system size $N$. This includes a 1-dimensional (1D) and 2-dimensional (2D) mixed-field Ising model (MFIM) with relatively small transverse field, and the 1D longitudinal-field Heisenberg XXZ model. We also show the effectiveness of our strategy in tracking different observables, including magnetization, spin current auto-correlation functions, and out-of-time-order correlators (OTOCs). We summarize the advantages achieved using our approaches in Tab.~\ref{Tab:SavingsSummary}.

\begin{table}[]
\renewcommand{\arraystretch}{1.5}
\begin{tabular}{|l|l|l|l|l|}
\hline
\textbf{Model}   & \textbf{Observable}    & $n (N)$ (\textbf{P}) & $n (N)$ (\textbf{K}) & $n (N)$ (\textbf{H}) \\ \hline
1D MFIM & Magnetization & 10 (12)       & 7 (8)          & 7 (12)         \\ \hline
2D MFIM & Magnetization & 14 (16)       & -              & 7 (16)         \\ \hline
LFXXZ   & SCACF         & 10 (12)       & -              & -              \\ \hline
LFXXZ   & OTOC          & -             & 10 (10)        & 10 (10)        \\ \hline
\end{tabular}
\renewcommand{\arraystretch}{1}
\caption{Summary of the savings obtained using approximate shadow Hamiltonian simulation. $n(N)$ corresponds to size of shadow register (size of physical register). The letters $\textbf{P},\textbf{K},\textbf{H}$ respectively stand for the predefined basis, Krylov basis, and hybrid schemes.}
\label{Tab:SavingsSummary}
\end{table}

{\it 1D Mixed-field Ising model ---} In 1D, the MFIM Hamiltonian with a small transverse field is given by
\begin{equation}
    H_{\text{MFIM}} = J\sum_{i=1}^{N-1} X_i X_{i+1} + h_x\sum_{i=1}^N X_i + h_z\sum_{i=1}^N Z_i,
\end{equation}
with $h_z < J, h_x$. Throughout this article, we have chosen $J = h_x=1$ unless explicitly specified otherwise. The efficiency of the two main pruning methods is shown in Fig.~\ref{fig:MFIM_main_text}. The pruning threshold for the predefined basis is chosen to be $h_z^M$, as the multiplicative weights of the graph nodes take this form. On the other hand, for the Krylov basis method, due to orthonormalization, the threshold does not have a specific form and we choose the cutoff of the form $10^{-m}$ without loss of generality.

Panel (a) in Fig.~\ref{fig:MFIM_main_text} displays the qubit register size required for the pruned shadow Hamiltonian simulation as a function of the physical system size. For each pruning threshold considered, the shadow register size reaches a plateau as the system size increases. Consequently, for a fixed evolution time and desired accuracy, the dynamics of observables in arbitrarily large systems can be tracked using a fixed-size shadow register. In Supplementary Information~\ref{app:extra-MFIM}, we show that for a fixed threshold, once the register size saturates, increasing the physical system size does not degrade the accuracy of the time evolution. We also explore the efficiency of the pruning schemes for different observables and the strength of the magnetic field $h_z$ in the 1D MFIM in Supplementary Information~\ref{app:extra-MFIM}. As discussed before, the scalability of our method is limited only by the classical memory required to store the basis vectors in $\mathbf{S}_{\bar{O}}$. If the classical preprocessing step reaches saturation for a given observable, target accuracy, and evolution time, then the method can be applied to physical systems of arbitrarily large size, with no additional increase in computational cost.

In the predefined Pauli-basis pruning scheme, an observable such as average magnetization, being a linear combination of Pauli operators, allows the shadow simulation to be parallelized over $k$ registers, where $k$ is the number of Pauli terms in the observable.

Panel (b) of Fig.~\ref{fig:MFIM_main_text} shows that the accuracy of the pruned shadow Hamiltonian simulation improves as the pruning threshold is reduced, particularly at longer evolution times. The scaling of the threshold required to maintain a given accuracy with increasing time is discussed in Methods~\ref{app:perturbative-pruning}. We further show the absolute simulation error with respect to exact diagonalization in the Appendix~\ref{app:error-magnetization}. We note that the model system size $N=12$ was chosen because it already lies in the regime where the shadow register requires fewer qubits than the original system, demonstrating the onset of quantum resource savings, which increases progressively as the physical system size increases further. The same shadow register size can also be used to simulate a 100-qubit system up to the same evolution time while maintaining comparable accuracy. However, exact diagonalization is computationally intractable at this system size, preventing direct benchmarking of the results. For larger values of the transverse field $h_z$, maintaining a fixed threshold $h_z^m$ leads to a loss of accuracy in the magnetization because the neglected contributions remain significant. While increasing $m$ improves the accuracy by retaining more operators, it also enlarges the cardinality of the truncated set $\mathbf{S}_{\bar{O}}$, thereby increasing the classical preprocessing cost. In practice, this becomes computationally challenging once $|\mathbf{S}_{\bar{O}}| \sim \mathcal{O}(2^{25})$. For example, the 1D MFIM model with $12$ physical spins, a choice of $h_z=0.4$ makes the required size of the shadow register larger than the system. The size of the shadow register would saturate at a larger value of $N$ for this set of model parameters. In the extreme case where $h_z\to1.0$, the full Pauli algebra would be important, making the approximate shadow Hamiltonian simulation approach not effective.  Consequently, the method is most effective when the transverse field $h_z$ is not too large compared to the characteristic interaction scale $J$ of the system.

Compared to the predefined-basis approach, the Krylov-basis pruning scheme exhibits a much slower growth of the operator set $\mathbf{S}_{\bar{O}}$ with increasing system size as shown in Fig.~\ref{fig:MFIM_main_text}(c). For a fixed target accuracy, the shadow register size in the Krylov-basis approach saturates at smaller system sizes than in the predefined-basis method, e.g., for $\epsilon = h_z^3$, shadow register size $n$ saturates at 7 qubits for Krylov basis compared to 10 qubits for predefined basis. However, expressing the Krylov basis for the 1D MFIM in the Pauli basis leads to storing the coefficients of $\mathcal{O}(4^N)$ distinct Pauli strings for the same target accuracy. As discussed earlier, this also increases the cost of shadow state initialization since expectation values of all relevant Pauli strings on the initial state must be evaluated to construct the initial shadow state. Consequently, the direct Krylov-basis approach does not remain scalable with system size for the systems considered in this work when using the Pauli basis to represent the Krylov vectors. For this reason, the magnetization dynamics shown in Fig.~\ref{fig:MFIM_main_text} (d) is demonstrated for a system size $N=8$. 

\begin{figure*}[ht!]
    \centering
\includegraphics[width=1.0\linewidth]{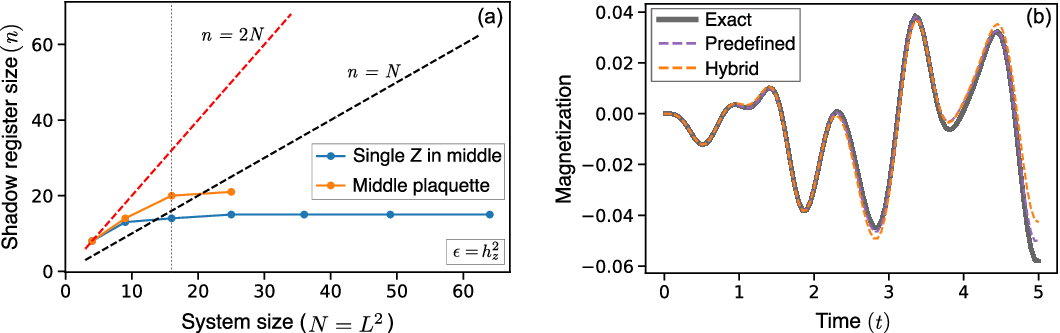}
    \caption{Pruning methods applied to the 2D MFIM with $h_z = 0.2$ and $\epsilon=h_z^2$. (a) The scaling of shadow register size with system size is shown for two different root observables (single-weight $Z$ in the middle site index and middle plaquette observable $Z_{x,y} Z_{x+1,y}Z_{x+1,y+1}Z_{x,y+1}$) of interest at a fixed pruning threshold $\epsilon = h_z^2$. (b) Average magnetization for a 2D MFIM for a $4\times 4$ (16 qubits) square grid. The dynamics of magnetization is calculated for an initial state $\ket{0}^{\otimes N/2}\otimes\ket{1}^{\otimes N/2}$. The shadow time evolution can be simulated with 14 qubits (purple dashed line) using the predefined basis pruning method with parallelization, demonstrating a savings in quantum resources. Due to the exponential scaling of the pure Krylov basis vectors, results corresponding to this method are numerically hard to obtain and not shown in this figure. On the other hand, the hybrid method can be used to further reduce the required shadow register size to $7$ for the same $4\times 4$ spin system while maintaining comparable accuracy, as shown by the orange dashed line. Absolute error in magnetization dynamics for the pruning methods is shown in Supplementary Information~\ref{app:error-magnetization}.}
    \label{fig:2DMFIM-magnetization}
\end{figure*}

The hybrid approach overcomes this limitation by performing the Krylov-basis construction on the pruned operator set $\mathbf{S}_{\bar{O}}$, allowing us to simulate systems with $N \gg 8$. For the MFIM, the benefit of this strategy is particularly pronounced because the size of $\mathbf{S}_{\bar{O}}$ saturates with increasing system size. As a result, once this saturation regime is reached, each Krylov basis vector constructed from $\mathbf{S}_{\bar{O}}$ contains only a constant number of Pauli strings, independent of system size.
The reduction in the number of Pauli strings compared to the Krylov basis is shown more explicitly in the Supplementary Information~\ref{app:extra-MFIM}.
In addition to reducing the cost of generating the Krylov basis vectors, the hybrid method requires a shadow register that is smaller than that of the predefined-basis approach to achieve the same magnetization accuracy. For example, a twelve-spin system can be simulated using a seven-qubit shadow register, compared to ten qubits required by the predefined-basis method (Fig.~\ref{fig:MFIM_main_text} (e)).
Furthermore, Fig.~\ref{fig:MFIM_main_text} (f) shows that restricting the Krylov construction to the pruned set $\mathbf{S}_{\bar{O}}$ also improves the simulation accuracy. For a fixed Krylov pruning threshold, the hybrid approach yields a smaller magnetization error than the full Krylov method while simultaneously retaining favorable scaling with system size.

{\it 2D MFIM ---} To assess the pruning scheme beyond one-dimensional spin chains, we consider lattice systems on a square grid, a regime that is typically more challenging for standard tensor-network methods. Specifically, we study the MFIM with nearest-neighbor interactions defined on a square lattice. 
In Fig.~\ref{fig:2DMFIM-magnetization} (a),
we demonstrate the scaling of the shadow register size with increasing system size for different observables of interest. The system size $N$ is defined by the length $L$ of a square grid, where the number of spins is $N = L^2$. As $L$ is increased, the shadow set size for a single qubit Pauli operator $Z$ in the middle-site index saturates quite early. For single-qubit composite observables like the magnetization, the minimum required shadow register size is defined by this saturation value. On the other hand, for plaquette observables like $Z_{x,y} Z_{x+1,y}Z_{x+1,y+1}Z_{x,y+1}$, the saturation occurs at a larger system size, as shown in Fig.~\ref{fig:2DMFIM-magnetization} (a), where the central plaquette operator is chosen as the observable.

To assess the accuracy of the simulated dynamics, we study the time evolution of the magnetization, similar to the one-dimensional MFIM. As shown in Fig.~\ref{fig:2DMFIM-magnetization} (a), the magnetization dynamics of a $4\times4$ spin system can be accurately simulated using a 14-qubit shadow register with the pruned shadow Hamiltonian constructed using $\epsilon = h_z^2$, thereby reducing the quantum resources required relative to the original 16-spin system. The close agreement between the magnetization dynamics obtained from the pruned shadow simulation and those from exact diagonalization in Fig.~\ref{fig:2DMFIM-magnetization} (b) validates the accuracy of our approach. We further show the absolute simulation error with respect to exact diagonalization in the Appendix~\ref{app:error-magnetization}. Moreover, the saturation of the shadow register size $(n)$ shown in Fig.~\ref{fig:2DMFIM-magnetization} (a) suggests that, for a fixed evolution time and target accuracy, magnetization dynamics in arbitrarily large two-dimensional systems can be computed using a constant-size shadow register. Since exact diagonalization becomes computationally prohibitive for larger system sizes $N>20$, we limit our study to sizes where exact diagonalization remains tractable. The hybrid pruning method further reduces the required register size. In particular, by exploiting parallelization, the $4\times4$ spin system can be simulated using only 7 qubits (with a Krylov cutoff $\epsilon_2 = 10^{-10}$) while achieving an accuracy comparable to that of the predefined-basis method, as shown in Fig.~\ref{fig:2DMFIM-magnetization} (b).

{\it Longitudinal-field XXZ model ---} In 1D, the Hamiltonian for the LFXXZ model is given by:
\begin{multline}
    H_{\text{LFXXZ}} = J\sum_{i=1}^{N-1} \left(X_i X_{i+1} +  Y_i Y_{i+1}\right) + \\J_z\sum_{i=1}^{N-1} Z_i Z_{i+1} + h_z\sum_{i=1}^N Z_i,
\end{multline}
where $J_z,h_z\leq 1$. We consider two applications: spin transport ($h_z=0, J = 1/4$) and OTOCs ($J=h_z=1.0,J_z=0.3$). For the spin-transport example, we employ the predefined-basis pruning scheme to demonstrate the effectiveness of parallelizing the shadow simulation when the observable is a linear combination of many terms, including multi-qubit Pauli strings. In contrast, the OTOC example highlights the advantages of the Krylov-basis and hybrid approaches, where savings in qubit register size can be achieved even for relatively small system sizes. This example also illustrates the additional reductions in the memory overhead to store the Krylov basis vectors enabled by the hybrid pruning scheme.

\begin{figure}[ht!]
    \centering
    \includegraphics[width=1.0\linewidth]{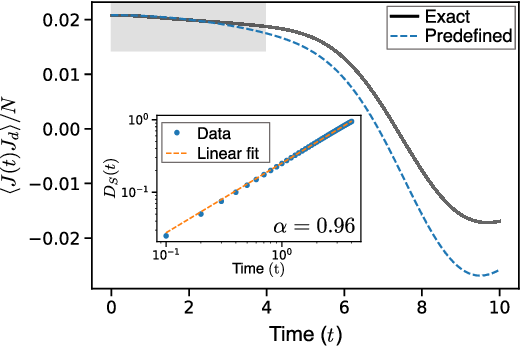}
    \caption{Study of a two-point correlator using pruned shadow simulation. The spin current autocorrelation function is plotted with time for a $12$-qubit Heisenberg XXZ model with $J_z = 0.05$ (ballistic regime). $J_d$ is the current across the domain wall and $J(t)$ is the total spin current. The shaded region shows the physically meaningful regime before boundary effects due to the finite system size become significant, which is also used to extract the scaling of the diffusion coefficient $D_S(t)$ (inset). The scaling $\alpha =0.96$ is close to the expected value of $\alpha=1$ in the ballistic regime. To perform the shadow simulation (with up to $10$ qubits), the predefined Pauli basis pruning scheme is used with a threshold of $J_z^2$.}
    \label{fig:spin-transport}
\end{figure}

\begin{figure}
    \centering
    \includegraphics[width=1.0\linewidth]{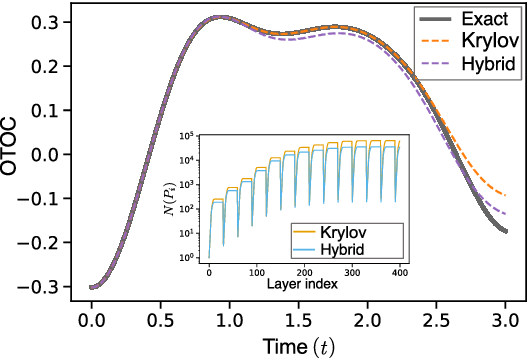}
    \caption{Study of a four-point correlator using the pruned shadow Hamiltonian method. The behavior of the OTOC ($W\equiv Z_1,V\equiv X_1$) with time is shown for a system size $N=5$ for the LFXXZ model with $J=h_z=1.0,J_z=0.3$ with the characteristic fast growth at early times. The cutoff used for the Krylov basis pruning scheme is $\epsilon_2=J_z^8$ with an additional Pauli pruning cutoff of $\epsilon_1=J_z^6$ for the hybrid scheme. The inset shows the number of Pauli strings, $N(P_i)$, appearing in the $i$-th operator of the Krylov basis. The two-step pruning procedure of the hybrid method results in lower number of Pauli strings contained in each operator, leading to a reduced cost for storing the Krylov vectors and shadow state initialization.}
    \label{fig:OTOC-vs-time}
\end{figure}

Spin transport in the one-dimensional XXZ Heisenberg model describes how local spin excitations, or spin current, spread through the chain under the Hamiltonian. A standard diagnostic of spin transport is the spin-current autocorrelation function (SCACF)
$C_J(t)= \frac{1}{N} \text{Re}\left[\sum_{i,j}\left\langle J_i(t) J_j(0) \right\rangle\right]$, where  $J_i = \left(
X_i Y_{i+1} - Y_i X_{i+1}\right)/4$ is the local current operator. Its late-time behavior distinguishes different transport regimes:
$J_z < 0.25$, $J_z = 0.25$, and $J_z>0.25$ indicating ballistic, superdiffusive, and diffusive transport, respectively \cite{JepsenNature2020,LeePRL2026}. Here, we focus on the ballistic regime given by $J_z<0.25$, where the SCACF decays very slowly. For a domain wall initial state $\ket{0}^{\otimes N/2}\otimes\ket{1}^{\otimes N/2}$, we study the SCACF $C_J(t)= \frac{1}{N} \text{Re[}\left\langle J(t) J_d(0) \right\rangle]$, where $J(t)$ is the total current operator and $J_d(0)$ is the local spin current across the domain wall for a 12-qubit system, using a shadow simulation that requires a maximum of $10$ qubits (allowing for parallelization) to calculate the correlation function (see Methods~\ref{app:twoptcorr} for more details).
Ignoring the finite-size effect for the small system, we can extract the time-dependent diffusion coefficient $D_S(t) = \int_0^t C_J(t')\,dt'$, which shows an asymptotic scaling $\sim t^\alpha$ with $\alpha = 1$ in the ballistic regime \cite{JepsenNature2020,LeePRL2026}. From our shadow simulation data, we obtain $\alpha = 0.96$, as expected in the ballistic regime. Figure~\ref{fig:spin-transport} shows the quality of the pruned shadow simulation compared to the exact time evolution along with the linear fit to obtain the diffusion coefficient.

We can also leverage the pruning scheme to calculate higher-order correlators as well. For example, the OTOC is an important four-point correlator used in the study and classification of quantum chaos in systems at finite temperature. Here, we consider the protocol introduced in Ref.~\cite{SundarNJoP2022} (see Methods~\ref{app:OTOC} for more details) to measure the OTOC given by:
\begin{equation}
    \text{OTOC}(t)=\frac{1}{Z}\Tr\left(e^{-\beta\,H/2}\,W^{\dagger}\,V^{\dagger}(t)\,W\,e^{-\beta\,H/2}\,V(t)\right),
\end{equation}
where $Z$ is the partition function and $\beta$ is the inverse temperature. $V(t)$ is a perturbation that is evolved by $H$ and $W$ detects the spread of $V(t)$. Here, we take $W=Z_1,V=X_1$. In Fig.~\ref{fig:OTOC-vs-time}, we numerically simulate the OTOC of the $5$-qubit LFXXZ model using the Krylov and hybrid pruning schemes. For the Krylov-basis pruning schemes, we take the cutoff $\epsilon_1=J_z^8$, along with a secondary Pauli-basis cutoff of $\epsilon_2=J_z^6$ for the hybrid scheme. For chaotic systems, the OTOC is expected to show a sharp exponential rise at early times, where the rate of the rise quantifies the quantum chaos in the form of a Lyapunov exponent. After the operator has scrambled to the available degrees of freedom, the OTOC saturates and oscillates. Both of these key features are captured well by the approximate shadow Hamiltonian simulation in Fig.~\ref{fig:OTOC-vs-time}. Finally, for small finite systems like the one considered here, the OTOC can decrease at late times. This is because the system is too small to observe true thermalization. The scaling of the qubit-register size required to time-evolve the OTOC using approximate shadow Hamiltonian simulation  versus the size of the physical system is presented in Methods~\ref{app:OTOC}, together with a review of the measurement protocol and its adaptation to shadow Hamiltonian simulation.

Through the above examples, we have shown that for many physically interesting models with characteristic small parameters, pruning can be very efficient for calculating the evolution of observable expectation values, autocorrelation functions, and higher-order correlators. However, the effectiveness of the truncation scheme, even in the presence of a small parameter, depends on the model and the choice of observable(s), as demonstrated in Supplementary Information~\ref{app:negative-example}. We also provide an example of a model with a non-local all-to-all connected Hamiltonian where pruning can be efficient in Supplementary Information~\ref{app:PLIM}.


\section*{Discussion}
In this article, we presented an algorithm to perform a controlled approximation of time-evolved observable expectation values under interacting many-body time-independent Hamiltonians. We removed the primary bottleneck of the shadow Hamiltonian simulation method, allowing the simulation of the dynamics of interacting systems without incurring an expensive qubit overhead to encode the shadow state. The encoding we used can be viewed as the spin-vector representation on an approximately closed algebra.

We focused on three manifestations of this approximation technique to identify the most important elements of the algebra for a set of target observables. The first method prunes a predefined operator basis by building a graph with the target observables as the root vertices. The second approach uses an operator Lanczos algorithm to build an approximate Krylov basis starting again with the operator(s) of interest. We further proposed a hybrid method that combines the strengths of both methods to create a Krylov basis constructed on the pruned set of predefined basis vectors. This reduces the required shadow register size for encoding the shadow Hamiltonian compared to the predefined basis method and makes the Krylov method scalable with system size. We benchmarked our algorithm using an array of numerical examples of lattice models in one and two dimensions, including one- and higher-point correlators. We note that there could be other pruning schemes that realize approximate shadow Hamiltonian simulation that are more appropriate for other applications. Possible extensions of pruned shadow Hamiltonian simulation to chemistry and high-energy physics Hamiltonians with small parameters are left for future work. Such extensions will also require identifying basis sets that are well suited to the corresponding pruning schemes, such as fermionic, Majorana, or Fock bases. It may also be worth considering making the generation of the pruned shadow set a fully quantum routine, which would make the algorithm more scalable, perhaps in a fault-tolerant setting. We leave this exploration for future work.

As a quantum algorithm, the efficiency of the method depends on operator spreading and can be compared with quantum simulation approaches using light-cone propagation. However, in light-cone methods, the effective velocity of the operator front either needs to calibrated from dynamics simulations classically or a Lieb-Robinson bound \cite{LiebCommPhys1972,WangPRXQ2020} must be used, which usually results in a larger required register size than needed. On the other hand, in our pruning schemes, the small parameters in the Hamiltonian play the central role in defining the required register size and provide a systematic way of building the shadow Hamiltonian without dynamic calibration. Here, the classical pre-processing step includes creating the basis vectors, which is usually much cheaper than performing a classical simulation of time evolution. In the Krylov basis and hybrid pruning scheme, the parent operator is allowed to be a composite operator and other basis operators may include different linear combinations of non-local operators. This is usually much more efficient in terms of register size compared to light-cone propagation methods, where the light cone of each individual operator needs to be taken into account separately. The systematic construction of the pruned predefined basis set and the Krylov basis also allow us to combine their advantages into the hybrid method to further reduce the quantum resources needed for the shadow simulation. Additionally, shadow Hamiltonian simulation provides simultaneous access to time-evolved expectation values of other operators, which may be of interest. 

As discussed above, our approximate shadow-simulation algorithm can also serve as a classical simulation technique, depending on the dimension and structure of the pruned shadow Hamiltonian. State-of-the-art classical techniques such as tensor networks and Pauli propagation take advantage of the structure of the Hamiltonian (locality, conserved quantities, etc.) to go beyond the reach of exact diagonalization. If the shadow Hamiltonian has these features, then our approach can be combined with these existing classical methods to improve their efficacy. A systematic analysis is needed to assess the impact of combining these approaches. Our approximate shadow-simulation algorithm may also offer advantages over existing methods; for example, unlike tensor-network-based approaches, it is easily generalizable to 2D and 3D models. 

For the results in this work, we have used exact diagonalization, and remained agnostic to the way the dynamics of the shadow Hamiltonian would be implemented on a quantum hardware. There are different aspects to realizing a shadow Hamiltonian simulation on a quantum computer that require further study. For example, the dimensionality of the shadow Hamiltonian need not be of the form $2^n$, where $n\in\mathbb{Z}_+$ as it is determined by the number of operators in the pruned set $\mathbf{S}_{\bar{O}}$. Simulating the shadow Hamiltonian on a qubit-based quantum computer may thus require block encoding, or one could instead consider a pruning scheme where a fixed number of operators is kept in the algebra, instead of a numerical cutoff on the weight of the operators.

Another important direction for future work is to study the structure of approximate shadow Hamiltonians that emerge for various types of problems and to estimate the resource requirements for the corresponding quantum simulation. This requires a careful analysis of the locality of the Hamiltonian, gate complexity, the complexity of preparing the initial shadow state, and an optimal time-evolution algorithm suitable for the shadow Hamiltonian. This would depend not only on the target model and observables, but also on the choice of pruning scheme and the pre-defined basis.


\begin{acknowledgments}
This work was supported by the U.S. Department of Energy, Office of Science, Office of Advanced Scientific Computing Research under Award Number DE-SC0025430.
\end{acknowledgments}


\section*{Author Contributions Statement}
E. Barnes conceived the project. S. E. Economou contributed in refining the algorithm. A. Chakraborty and B. Sambasivam developed the pruning schemes, built the codebase, performed the numerics, and derived the error bounds. K. Shirali, H. Nelson, and M. Ram\^{o}a contributed to the conceptual improvement of the schemes. All authors contributed to the writing of the manuscript. \\~\\


\section*{Competing Interests Statement}
The authors declare no conflicts of interest.\\


\renewcommand{\thesection}{}
\renewcommand{\thesubsection}{\Alph{subsection}}
\section*{Methods} 
\label{methods}
In the Methods section, we provide the pseudocodes for our pruning schemes, additional information required to perform the simulations for two-point correlators and OTOCs in Sec.~\ref{sec:Results}, and derive the error bounds for the pruned shadow Hamiltonian simulation algorithms introduced in the main text.


\subsection{Pseudocodes}\label{app:pseudocodes}
The pseudocodes for the pruning schemes used in Sec.~\ref{sec:predefined-basis} and Sec.~\ref{sec:constructed-basis} are shown in Algorithms~\ref{alg:graph-pruning} and \ref{alg:pruned-krylov-subspace}. For both code segments, $H$ denotes the Hamiltonian, our observable of interest is $\bar{O}$, the threshold is $\epsilon$, and the maximum number of layers in the graph is $L_{\rm max}$. For the first scheme shown in Algorithm~\ref{alg:graph-pruning} and the hybrid version of Algorithm~\ref{alg:pruned-krylov-subspace}, the predefined basis is indicated by $\mathbf{B}$ and $\bar{O}$ is an element of $\mathbf{B}$. The accumulated weight of an operator $\Omega(O)$ is defined in Eq.~\eqref{eq:AccumWts}.

\begin{figure}[ht!]
\begin{algorithm}[H]
\caption{Pruning of operator-growth graph}
\label{alg:graph-pruning}
\begin{algorithmic}[1]
\vspace{0.5em}
\Statex\hspace{-\algorithmicindent}\textbf{Input:} $H, \bar{O}, \mathbf{B},\epsilon, L_{\rm max}$.

\Statex\hspace{-\algorithmicindent}\textbf{Output:} Pruned shadow set 
$\mathbf{S}_{\bar{O}}$
\vspace{0.5em}
\State Initialize $\mathbf{S}_{\bar{O}}=\{\bar{O}\}$ and $\Omega(\bar{O})=1$

\For{$\ell=0,\dots,L_{\max}-2$}

    \State Initialize $\mathcal{N}_\ell=\emptyset$

    \ForAll{$O\in \mathbf{S}_{\bar{O}}$ at layer $\ell$}

        \State Compute $[H,O]=\sum_{O'\in\mathbf{B}} c_{O'} O'$

        \ForAll{generated operators $O'$}

            \State $\widetilde{\Omega}=\Omega(O)\,|c_{O'}|$

            \If{$\widetilde{\Omega}\ge \epsilon$}

                \If{$O'\notin \mathbf{S}_{\bar{O}}$
                \textbf{or}
                $\widetilde{\Omega}>\Omega(O')$}

                    \State Add $O'$ to $\mathcal{N}_\ell$
                    \State Update $\Omega(O')\gets \widetilde{\Omega}$

                \EndIf
            \EndIf

        \EndFor
    \EndFor

    \If{$\mathcal{N}_\ell=\emptyset$}
        \State \textbf{break}
    \EndIf

    \State Update
    $
    \mathbf{S}_{\bar{O}}
    \gets
    \mathbf{S}_{\bar{O}} \cup \mathcal{N}_\ell
    $

\EndFor

\State \Return $\mathbf{S}_{\bar{O}}$

\end{algorithmic}
\end{algorithm}
\end{figure}

In Algorithm~\ref{alg:pruned-krylov-subspace}, we present the Krylov pruning scheme. The hybrid scheme can be implemented by first building a subspace $\mathbf{S}$ using Algorithm~\ref{alg:graph-pruning} and projecting onto $\mathbf{S}$ the new operator obtained at each iteration by the action of $\mathcal{A}_H$. Subsequently, a standard orthonormalization procedure is employed to build a Krylov subspace within $\mathbf{S}$, making the algorithm more scalable with the size of the system.

\begin{figure}[ht!]
\begin{algorithm}[H]
\caption{Pruned Krylov subspace construction}
\label{alg:pruned-krylov-subspace}
\begin{algorithmic}[1]
\vspace{0.5em}
\Statex \hspace{-\algorithmicindent} \textbf{Input:} $H, \bar{O}, \epsilon, L_{\rm max}$, {\rm tol}, $\epsilon'$ (for hybrid), $L'_{\text{max}}$ (for hybrid),$\mathbf{B}$ (for hybrid).
\Statex \hspace{-\algorithmicindent} \textbf{Output:} Pruned Krylov basis $\mathcal{K}$.
\vspace{0.5em}
\State $\mathbf{S} \gets$ \textsc{Algorithm 1}($H,\bar{O},\mathbf{B},\epsilon',L'_{\text{max}}$) (for hybrid)
\State $K_0 \gets \bar{O}/\|\bar{O}\|$
\State $\mathcal{K} \gets \{K_0\}$
\State $w \gets 1$
\For{$m = 0,\ldots,L_{\max}-2$}
    \State $\widetilde{O}_{m+1} \gets [H,K_m]/i$
    \State $\widetilde{O}_{m+1} \gets$ \textsc{Project}($\mathbf{\widetilde{O}_{m+1},S}$) (for hybrid) 
    \State $n_{\mathrm{pre}} \gets \|\widetilde{O}_{m+1}\|$
    \State $O_{m+1}^{\perp} \gets \widetilde{O}_{m+1}
    - \sum_{r=0}^{m} \langle K_r,\widetilde{O}_{m+1}\rangle K_r$
    \State $n_{\mathrm{post}} \gets \|O_{m+1}^{\perp}\|$
    \If{$n_{\mathrm{post}} < {\rm tol}$}
        \State \textbf{break}
    \EndIf
    \State $w \gets w\, n_{\mathrm{post}}/n_{\mathrm{pre}}$
    \If{$w < \epsilon$}
        \State \textbf{break}
    \EndIf
    \State $K_{m+1} \gets O_{m+1}^{\perp}/n_{\mathrm{post}}$
    \State $\mathcal{K} \gets \mathcal{K} \cup \{K_{m+1}\}$
\EndFor
\State \textbf{return} $\mathcal{K}$
\end{algorithmic}
\end{algorithm}
\end{figure}


\subsection{Measurement of OTOCs in the shadow picture}\label{app:OTOC}
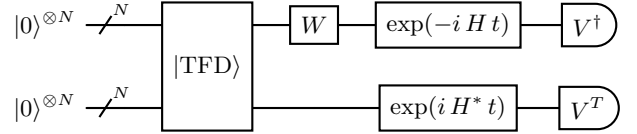
\begin{figure}
    \centering
    \begin{quantikz}[classical gap=0.1cm]
        \lstick{$\ket{0}^{\otimes N}$} & \qwbundle{N} & \gate[2]{\ket{\text{TFD}}}& \gate{W} & \gate{\exp(-i\,H\,t)} & \meterD{V^{\dagger}}\\
        \lstick{$\ket{0}^{\otimes N}$} & \qwbundle{N} & & & \gate{\exp(i\,H^*\,t)} & \meterD{V^{T}}
    \end{quantikz}
    \caption{Protocol to measure the OTOC introduced in Ref.~\cite{SundarNJoP2022}. The initial state is the TFD state on $2N$ qubits.}
    \label{fig:OTOCcirc}
\end{figure}

Here, we review the protocol introduced in Ref.~\cite{SundarNJoP2022} to measure the finite temperature OTOC:
\begin{equation}
    \text{OTOC}(t)=\frac{1}{Z}\Tr\left(e^{-\beta\,H/2}\,W^{\dagger}\,V^{\dagger}(t)\,W\,e^{-\beta\,H/2}\,V(t)\right).
\end{equation}
The quantum circuit to measure this quantity is given in Fig.~\ref{fig:OTOCcirc}. The initial state is the thermofield double (TFD) state defined as
\begin{equation}
    \ket{\text{TFD}}\equiv\frac{1}{\sqrt{Z}}\sum_{k}e^{-\beta\,E_k/2}\ket{\psi_k}\otimes\ket{\psi^*_k},
\end{equation}
where $E_k,\ket{\psi_k}$ are the eigenvalues and eigenstates of $H$. Note that for a $N$-qubit Hamiltonian, the TFD state is a pure state on $2N$ qubits.

\begin{figure}
    \centering
    \includegraphics[width=1.0\linewidth]{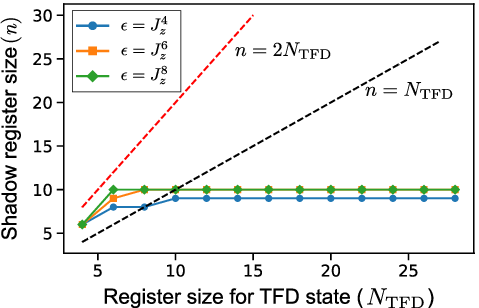}
    \caption{Scaling of the number of qubits required for pruned shadow Hamiltonian simulation of the thermal OTOC for the LFXXZ model with $J_x=J_y=h_z=1.0,J_z=0.3,\beta=2.0$ vs the number of physical qubits in the TFD state. The different colored points correspond to various cutoffs for the predefined Pauli basis pruning scheme. The red dotted line is the qubit requirements for full shadow Hamiltonian simulation of the OTOC measurement in Fig.~\ref{fig:OTOCcirc}, and the black dotted line is the scaling of the number of physical qubits in $\ket{\text{TFD}}$.} 
    \label{fig:OTOC_scaling}
\end{figure}

This protocol is straightforwardly amenable to a Hamiltonian simulation/observable evolution problem. The initial state in this case is
\begin{equation}
    \ket{\Psi_0}\equiv W\ket{\text{TFD}},
\end{equation}
the Hamiltonian is
\begin{equation}
    H_D\equiv H\otimes\mathds{1}-\mathds{1}\otimes H^*, 
\end{equation}
and the observable being measured is
\begin{equation}
    \bar{O}=V^{\dagger}\otimes V^T.
\end{equation}
Our pruned shadow Hamiltonian simulation algorithm can be applied directly to this problem using $\bar{O}$ as the root observable. Given the structure of the Hamiltonian $H_D$ and the target observable $\bar{O}$, the algebra has a product form:
\begin{equation}
    \mathfrak{g}(H_D,\bar{O})\equiv\mathfrak{g}(H,V^{\dagger})\times\mathfrak{g}(H^*,V^T),
\end{equation}
which makes the computation of the pruned algebra efficient. In Fig.~\ref{fig:OTOC_scaling}, we plot the scaling of the number of qubits required to apply our algorithm to measure the OTOC using the circuit shown in Fig.~\ref{fig:OTOCcirc} for the LFXXZ model with $J_x=J_y=h_z=1.0,J_z=0.3,\beta=2.0$. The dotted black line shows the scaling of the number of qubits in the TFD state, and any point below that line corresponds to an advantage in the number of qubits needed to encode the shadow state and shadow Hamiltonian. 

\subsection{Two-point correlators in the shadow picture}\label{app:twoptcorr}
In the original shadow simulation formulation~\cite{SommaNatComm2025}, the authors also introduced a way to determine the expectation values of unequal-time correlators \cite{SommaNatComm2025}. More specifically, for a two-point correlator $O(t) O(t')$, the shadow simulation method can be used  by defining the initial shadow state 
\begin{equation}
    \ket{\rho(0,0);\mathbf{S}} := \frac{1}{\sqrt{A}} \begin{pmatrix}
        \langle O_1(0) O_1(0) \rangle \\
        \vdots\\
        \langle O_1(0) O_K(0) \rangle\\
        \langle O_2(0) O_1(0) \rangle\\
        \vdots\\
        \langle O_K(0) O_K(0) \rangle  \\
    \end{pmatrix}
\end{equation}
on a $2N$ qubit register where $N$ is the register size needed to encode the shadow Hamiltonian $H_{\mathbf{S}}$. $K$ is the size of the set $\mathbf{S} = \{O_1,\dots,O_K\}$ and the normalization of the shadow state is given by the quantity $A$. This shadow state then evolves according to the following equation
\begin{align}
\frac{\partial}{\partial t}\ket{\rho(t,t'); \mathbf{S}}
&=
-\mathrm{i}\left(H_S \otimes \mathds{1}_K\right)\ket{\rho(t,t'); \mathbf{S}} \, ,
\\
\frac{\partial}{\partial t'}\ket{\rho(t,t'); \mathbf{S}}
&=
-\mathrm{i}\left(\mathds{1}_K \otimes H_S\right)\ket{\rho(t,t'); \mathbf{S}} \, .
\end{align}

The two-point correlators are then elements of the evolved shadow state $\ket{\rho(t,t');\mathbf{S}}$. 

For the spin-current autocorrelation function, $t'$ is fixed at $t'=0$. So, we do not need to solve the second partial differential equation. Thus, we only need to solve the Schr\"{o}dinger equation
\begin{equation}
    \frac{d}{d t}\ket{\rho(t,0); \mathbf{S}}=
-\mathrm{i}\left(H_S \otimes \mathds{1}_K\right)\ket{\rho(t,0); \mathbf{S}}. 
\end{equation}

To define the shadow state, we need to start with a set $\mathbf{S}$ that includes local spin current $J_i$ and the domain wall spin current $J_d$, such that the shadow state has an element $J_i(0) J_d(0)$ at initial time. This can be achieved by taking the union of the pruned sets $\mathbf{S}_{J_i} \cup \mathbf{S}_{J_d}$. For a system size of twelve qubits considered in the main text and the threshold choice, this union requires a shadow Hamiltonian encoding with a maximum of five qubits. Subsequently, the corresponding shadow state encoding requires a maximum of ten qubits. To obtain the spin current autocorrelation function, a parallel simulation can be run for all $N-1$ different values of $i$.


\subsection{Error bound for predefined basis pruning}\label{app:perturbative-pruning}
Consider a Hamiltonian of the form $H=H_0+\xi\,H_1$, where $H_0$ is an exactly solvable portion, and $H_1$ is a perturbation of strength $\xi$. We are going to derive a bound on the error in the pruned evolution of an anchor observable $O$ under $H$ using \textsc{Algorithm} 1. In the interaction picture with respect to $H_0$, any operator $A$ is $A_I(t) = e^{iH_0\,t}Ae^{-iH_0\,t}$. The exact time evolution of the observable can be written as a Dyson series:
\begin{multline}
    O(t) = O_I(t)+\sum_{n=1}^{\infty}(i\xi)^n\int_0^t\,dt_1\int_0^{t_1}dt_2\\\cdots\int_0^{t_{n-1}}dt_n\left[H_{1I}(t_n),\left[\cdots\left[H_{1I}(t_1),O_I(t)\right]\right]\right].
\end{multline}
The norm of the $k^{\text{th}}$-order term in $\xi$ is upper bounded by:
\begin{multline}
    \vert\vert O^{(k)}(t)\vert\vert\leq\xi^k\int_0^t dt_1\int_0^{t_1}\cdots\int_0^{t_{k-1}}dt_k\left(2^k\vert\vert H_1\vert\vert^k\vert\vert O\vert\vert\right)\\
    =\frac{\vert\vert O\vert\vert}{k!}\left(2\xi\vert\vert H_1\vert\vert t\right)^k.
\end{multline}
Keeping terms up to $\mathcal{O}(\xi^M)$, the total error is the residue
\begin{equation}
\vert\vert R_M(t)\vert\vert=\vert\vert O(t)-\sum_{j=0}^MO^{(j)}(t)\vert\vert\leq\vert\vert O\vert\vert\sum_{k=M+1}^{\infty}\frac{x^k}{k!},
\end{equation}
where $x=2\xi\vert\vert H_1\vert\vert t$. This residual can be upper bounded using Taylor's theorem, yielding the error bound:
\begin{equation}\label{eq:PertResidueBound}
    \vert\vert R_M(t)\vert\vert\leq\vert\vert O\vert\vert \frac{(2\xi\vert\vert H_1\vert\vert t)^{M+1}}{(M+1)!}\exp\left(2\xi\vert\vert H_1\vert\vert t\right).
\end{equation}
Going forward, we will assume without loss of generality that the anchor operator is normalized, i.e., $\vert\vert O\vert\vert=1$. Using Eq.~\eqref{eq:PertResidueBound}, given a target precision $\delta$, we will derive an asymptotic scaling of the required power $M$ of $\xi$ to choose as the pruning threshold in Algorithm 1. Rearranging the condition $\vert\vert R_M(t)\vert\vert\leq\delta$ and using Stirling's approximation, we obtain the following:
\begin{equation}
    y\ln y\geq\frac{1}{e}\left(1-\frac{1}{x}\ln\delta\right),
\end{equation}
where $y=(M+1)/(ex)$. Next, we consider two limiting regimes of the r.h.s and derive the scaling of $M$:
\begin{itemize}
    \item Time-dominated regime ($x\gg\ln\delta^{-1}$):
    $M\ln\frac{M}{ex}\geq x$. A scaling of $M=cx$ satisfies this equality provided $c>e^2$. Therefore, the asymptotic scaling in this limit is $M=\mathcal{O}(2\xi\vert\vert H_1\vert\vert t)$.
    \item Precision-dominated regime ($\ln\delta^{-1}\gg x$):
    $M\ln\left(\frac{M}{ex}\right)\geq\ln\delta^{-1}$. A scaling of $M=\frac{2L}{\ln L}$ satisfies this inequality, yielding the asymptotic scaling $M=\mathcal{O}\left(\frac{\ln\delta^{-1}}{\ln\ln\delta^{-1}}\right)$.
\end{itemize}
Putting these scalings together, we get the overall asymptotic scaling of $M$ valid in all regimes of $x$ and $\delta$:
\begin{equation}
    M=\mathcal{O}\left(2\xi\vert\vert H_1\vert\vert t+\frac{\ln\delta^{-1}}{\ln\ln\delta^{-1}}\right).\label{eq:graph-errorbound-M}
\end{equation}

\begin{figure}
    \centering
    \includegraphics[width=0.8\linewidth]{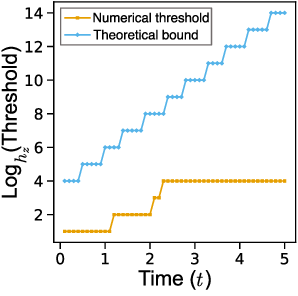}
    \caption{Threshold required to achieve a fixed accuracy $\delta=0.01$ as a function of time for an $N=12$ spin system compared with the analytical bound derived in Methods~\ref{app:perturbative-pruning}. The observable used for the error analysis is the single-qubit $Z$ operator acting on the central qubit for the MFIM in 1D.}
    \label{fig:error-bound-graph}
\end{figure}

The bound derived in Eq.~(\ref{eq:graph-errorbound-M}) is not tight, as demonstrated in Fig.~\ref{fig:error-bound-graph}, which shows the numerically obtained value of $M$ required to achieve a target accuracy $\delta$ for the one-dimensional MFIM.

\subsection{Error bound for Krylov basis pruning}\label{app:krylov-pruning}
An error bound for the Krylov-based pruning method can be derived using the Arnoldi residual estimate \cite{WangSIAM2017,MichelCompPhysComm2022}. Here we provide the essential steps for proving the bound.

Consider a pruned Krylov subspace of length $m$, defined by the matrix $V_m = (q_0,q_1,\dots,q_{m-1})$, where $q_i$ are a set of orthonormal vectors obtained through the process of building Krylov basis. $P_m = V_m V_m^\dagger$ defines a projection operator into the Krylov subspace. In our case, $q_0 = \bar{O}/||\bar{O}||$ is the normalized observable that we want to measure as a function of time. We define the Liouvillian superoperator $\mathcal{L}$ as $\mathcal{L}[O] = i[H,O]$. So, the Krylov basis is built using the repeated application of $\mathcal{L}$ on the root observable $\bar{O}$ and using a orthonormalization procedure like the Gram-Schmidt process. The Arnoldi iteration relation for Krylov subspace is then given by \cite{SaadSIAM2011}
\begin{equation}
    \mathcal{L}V_m = V_m A_m + \beta_m q_m e_m^\dagger \;, \label{eq:arnoldi-iteration}
\end{equation}
where $e_m = (0,0,\dots,0,1)$ is the $m$-th computational basis vector, $\beta_m$ is the norm of the orthogonal part outside the $m$-dimensional Krylov space, i.e. $\beta_m = || (I-P_m)\mathcal{L}q_{m-1}||$, and $A_m = V_m^\dagger \mathcal{L}V_m$. The elements of the matrix $A_m$ are defined as $(A_m)_{ij} = \langle q_i, \mathcal{L}q_j\rangle$, with $\langle\, \cdot\,,\,\cdot\,\rangle$ denoting the Hilbert-Schmidt inner product. Using Eq.~(\ref{eq:arnoldi-iteration}) and the Krylov subspace time evolution, it can be shown that the residual error $R_m(t) = ||\bar{O}|| \, (q_0(t) - q_0^{(m)}(t))$ is bounded by~\cite{WangSIAM2017, MichelCompPhysComm2022}
\begin{equation}
    ||R_m(t)|| \leq ||\bar{O}|| \beta_m \int_0^t |e_m^\dagger e^{sA_m}e_1|\; ds \;,\label{eq:first-bound-Krylov}
\end{equation}
where $q_0^{(m)}(t)$ is obtained from the time-evolution of $q_0$ performed within the pruned Krylov subspace. A derivation of Eq.~(\ref{eq:first-bound-Krylov}) is shown in Supplementary Information~\ref{SIsec:arnoldi-residue}.

However, our pruning criteria involves tracking the quantity $p_m = \prod_{j=1}^{m-1} \beta_j/\eta_j$, where $\eta_j = || \mathcal{L}q_{j-1}||$. In order to connect our pruning condition with the error bound derived above, we need to loosen the bound a bit further. The integrand in Eq.~(\ref{eq:first-bound-Krylov}) follows
\begin{equation}
    |e_m^\dagger e^{sA_m}e_1| \leq \sum_{n=0}^\infty \frac{s^n}{n!} \left|e_m^\dagger A_m^n e_1\right|\;.
\end{equation}
Due to the construction process, $e_m^\dagger(A_m)^ne_1 = 0$ for $n<m-1$. So, the sum reduces to
\begin{align}
     |e_m^\dagger e^{sA_m}e_1| &\leq \sum_{k=m-1}^\infty \frac{s^k}{k!} \left|e_m^\dagger A_m^k e_1\right| \nonumber\\
     & = \frac{s^{m-1}}{(m-1)!} \sum_{n=0}^\infty \frac{s^k (m-1)!}{(m-1+k)!} \left|e_m^\dagger A_m^{m-1+k} e_1\right| \label{eq:krylov-series-first-ineq}
\end{align}

To simplify, we note that in the leading term in the sum, $e_m^\dagger A_m^{m-1}e_1 = \prod_{j=1}^{m-1}\beta_j$. Any higher order terms will contain this product and can be bounded by 
\begin{equation}
   \left| e_m^\dagger A_m^{m-1+k} e_1 \right|\leq \left(\prod_{j=1}^{m-1} \beta_j \right) \Lambda^k
\end{equation}
where $||A_m|| \leq \Lambda$. Since $(m-1)!/(m-1+k)! \leq 1/k!$, the inequality in Eq.~(\ref{eq:krylov-series-first-ineq}) becomes
\begin{equation}
    |e_m^\dagger e^{sA_m}e_1| \leq \frac{s^{m-1}}{(m-1)!} \left(\prod_{j=1}^{m-1} \beta_j \right) e^{\Lambda s}\,.
\end{equation}
Noting $e^{\Lambda s}\leq e^{\Lambda t}$ for $s\leq t$, Eq.~(\ref{eq:first-bound-Krylov}) becomes
\begin{equation}
    ||R_m(t)|| \leq ||\bar{O}||  \left(\prod_{j=1}^{m} \beta_j \right) \frac{t^m}{m!} e^{\Lambda t}\;.
\end{equation}
We are now in a position to connect our pruning diagnostic quantity $p_k$ with the error bound
\begin{equation}
    ||R_m(t)|| \leq ||\bar{O}||\,  p_m\left(\prod_{j=1}^{m} \eta_j \right) \frac{t^m}{m!} e^{\Lambda t}\;. \label{eq:krylov-bound-aposteriori}
\end{equation}
For the pruning criteria $p_m<\epsilon$, we can reformulate the bound as
\begin{equation}
    ||R_m(t)|| \leq \epsilon\,||\bar{O}||\,    \frac{(\Lambda t)^m}{m!} e^{\Lambda t}\;. \label{eq;krylov-bound-final}
\end{equation}
where we have used $\eta_j \leq \Lambda$. $\Lambda$ can be chosen as the norm of the Liouvillian operator, which is upper bounded by $\Lambda \leq 2||H||$. This provides an a priori error bound for the Krylov pruning. For a local Hamiltonian and a local operator, the bound $\Lambda \leq 2||H||$ can be replaced by a tighter bound that depends on the operator support, rather than the system size $N$.

\begin{figure}[ht!]
    \centering
    \includegraphics[width=0.8\linewidth]{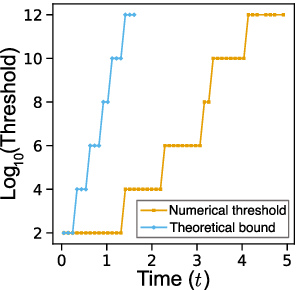}
    \caption{Theoretical and numerical threshold as a function of time using Krylov basis pruning for $N=8$ 1D MFIM with $h_z = 0.3$ with a desired accuracy $\delta = 0.01$. The theoretical threshold value is calculated using Eq.~(\ref{eq:krylov-bound-aposteriori}). } 
    \label{fig:error-bound-krylov}
\end{figure}

Similarly to the predefined basis pruning method, the error bound derived here is not a tight bound, as demonstrated in Fig.~\ref{fig:error-bound-krylov}. The threshold required to achieve a certain accuracy grows much slower with time compared to the threshold determined theoretically.

\subsection{Classical complexity of building $\mathbf{S}_{\bar{O}}$}\label{app:ClassComplex}
Here, we consider the classical computational complexity of building the shadow set $\mathbf{S}_{\bar{O}}$ from a root observable $\bar{O}$ using the pruning schemes introduced in the main text. 

\textit{Predefined basis---}Consider a Hamiltonian $H$ expanded in a predefined basis $\mathbf{B}$:
\begin{equation}
    H = \sum_{b\in\mathbf{B}}\beta_b \,b,
\end{equation}
where there are $N_H$ terms in the sum. Starting with an observable $\bar{O}\in\mathbf{B}$, the first non-root layer of the graph will have at most $N_H$ vertices. Similarly, the $l^{\text{th}}$ layer will have at most $N_H^l$ vertices. So, the asymptotic computational complexity of computing $n$ vertices of the graph has a worst-case scaling of $\mathcal{O}(N_H^l)$. Realistically, for problems like the ones considered in this work, the scaling will be milder than this because some subset of operators corresponding to vertices in adjacent layers would commute due to, for instance, disjoint support on qubits. This bound also works for any $\bar{O}$ and $H$, so additional constraints on the locality of both would ease the complexity. Most importantly, a non-zero threshold would eliminate a large portion of the vertices.

One such specialization was considered in the main text, where the Hamiltonian is of the form $H=H_0+\xi\,H_1$ with a cutoff of $\epsilon=\xi^M$. Let there be $N_{0,1}$ terms in the expansion of $H_{0,1}$ in the basis $\mathbf{B}$, where $N_0+N_1\equiv N_H$. Given the cutoff, in any layer of the graph, terms of $\mathcal{O}(\xi^j)$ where $j>M$ will not appear. As a result, the computational complexity of generating this graph scales as 
\begin{align}
    \mathcal{O}\left(\sum_{j=0}^M\binom{l}{j}N_0^{l-j}N_1^j\right).
\end{align}

\textit{Krylov basis---}For the practical implementation of the Krylov basis scheme, one still needs to pick a basis to write down the Hamiltonian and the target observables in---the ideal choice being one in which both are sparse. As before, let the Hamiltonian have $N_H$ terms in this basis and the root observable have $N_{\bar{O}}$ terms. The operator obtained from $n$ applications of the Liouvillian, $\mathcal{A}_H$ will contain atmost $\mathcal{O}(N_{\bar{O}}N_H^n)$ terms in this chosen basis. The orthongonalization procedure could, in principle, make this scaling worse. While the same savings from disjoint overlap of operators and locality of the Hamiltonian are present here as well, elements of the chosen basis with small coefficients are \emph{not} eliminated in the Krylov basis approach. Moreover, elements of the chosen basis that have already appeared in an operator can reappear in an operator at a later iteration. This can cause the number of terms in the chosen basis in each Krylov basis operator to explode, making it infeasible at scale for most problems of interest. In principle, every single element of the chosen basis could appear in the Krylov basis elements, leading to an intractable memory overhead.

\textit{Hybrid approach---}The hybrid approach overcomes this scaling issue by restricting the Krylov basis to a subspace of the full algebra. This is done by first identifying the important elements $\mathbf{S}$ of the chosen basis for a given observable $\bar{O}$ and then building a Krylov basis in the subspace spanned by these important elements. This allows for the crucial advantage in scaling of the predefined basis method to carry over here as well.

\putbib[biblio]
\end{bibunit}

\clearpage
\begin{bibunit}
\widetext
\begin{center}
\textbf{\large Supplementary Information}
\end{center}

\setcounter{section}{0}
\setcounter{equation}{0}
\setcounter{figure}{0}
\setcounter{table}{0}

\makeatletter
\renewcommand{\theequation}{S\arabic{equation}}
\renewcommand{\thefigure}{S\arabic{figure}}
\renewcommand{\thesection}{S.\Roman{section}}
\renewcommand{\bibnumfmt}[1]{[S#1]}
\renewcommand{\citenumfont}[1]{S#1}

\section{Derivation of Arnoldi residue \texorpdfstring{$R_m(t)$}{Rm(t)}}\label{SIsec:arnoldi-residue}

The Krylov approximation for the time evolution of the operator $q_0$ inside the Krylov subspace is given by $q_0^{(m)}(t) = V_m e^{tA_m}e_1$ \cite{SaadSIAM2011}. We now define the quantity $r_m(t)$ as
\begin{equation}
    r_m(t) = \frac{d}{dt}q_0^{(m)}(t) - \mathcal{L}q_0^{(m)}(t)\;.
\end{equation}
Using the expression for $q_0^{(m)}(t)$ we get $r_m(t) = -\beta_m q_m e_m^\dagger e^{tA_m} e_1 \nonumber$, so that
\begin{equation}
     ||r_m(t)|| = \beta_m |e_m^\dagger e^{tA_m}e_1|\;, \label{eq:instant-residue-krylov}
\end{equation}
where we have used $|q_m| = 1$ as $q_i$ are orthonormal vectors.

Now, the residual error in the Krylov time evolution can be defined as $R_m(t) = ||\bar{O}|| \, (q_0(t) - q_0^{(m)}(t))$. Defining $\varepsilon_m(t) = R_m(t)/||\bar{O}||$, we get
\begin{align}
    \dot{\varepsilon}_m(t) &= \dot{q}_0(t) - \dot{q}_0^{(m)}(t) \nonumber\\
    & = \mathcal{L}\varepsilon_m(t) - r_m(t) \nonumber\\
   \varepsilon_m(t) &= -\int_0^t e^{(t-s)\mathcal{L}} \,r_m(s) \, ds
\end{align}
Using Eq.~(\ref{eq:instant-residue-krylov}) we can then write the norm of the error residue as
\begin{equation}
    ||R_m(t)|| \leq ||\bar{O}|| \beta_m \int_0^t |e_m^\dagger e^{sA_m}e_1|\; ds \label{appeq:first-bound-Krylov}
\end{equation}
This is the error bound for the Krylov method, which can be numerically tracked a posteriori.

\section{Accuracy of other operators in the pruned shadow basis}

Here, we explore the accuracy in the evolution of operators that are not in the root layer of the graph in the predefined basis pruning scheme. In the case where the cutoff is zero (pure shadow Hamiltonian simulation), the evolution of any operator and their linear combinations can be studied exactly. In the presence of a non-zero cutoff $\epsilon$, there is a smaller set of observables that remain as accurate as those in the root layer of the graph. Consider, as before, a Hamiltonian of the form $H=H_0+\xi\,H_1$, where $\xi$ is a small parameter. From any given vertex, $V$ of the graph, there are two classes of edges that can emerge:
\begin{align}
    E_0&\equiv \left[\sigma,V\right],\quad\sigma\subset H_0\\
    E_1&\equiv \left[\sigma,V\right],\quad\sigma\subset H_1.
\end{align}
Throughout a path, from the root layer to $V$, any time an edge of type $E_1$ is encountered, the precision in the evolution of the operator corresponding to $V$ is reduced. Qualitatively, given a cutoff $\xi^M$ for the generation of the graph for some $M\in\mathbb{Z}_+$, the effective cutoff for the non-root vertex $V$ is $\xi^{M-N_{\xi}}$, where $N_\xi$ is the minimum number of edges of type $E_\xi$ encountered along any path connecting the root layer to $V$. 

\begin{figure}
    \centering
    \includegraphics[width=0.5\linewidth]{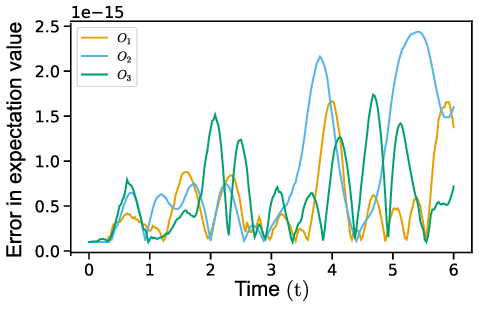}
    \caption{Absolute difference between pruned shadow Hamiltonian simulation of operators $O_1,O_2,O_3$, which are respectively $\mathcal{O}(1),\mathcal{O}(J_z),\mathcal{O}(J^2_z)$ in the $\epsilon=J_z^5$ graph generated using $\bar{O}$ as the root observable and simulation with the graphs generated using $O_1,O_2,O_3$ as the root observables with $\epsilon=J^5_z,J^4_z,J^3_z$ for the LFXXZ model with $J_x=J_y=h_z=1.0,J_z=0.3$.}
    \label{fig:error-by-layer}
\end{figure}

There can exist paths where only the $E_0$ class of edges is encountered starting from the root vertices. The time-evolution of any operator along such paths will be as accurate as those in the root layer. We demonstrate this with a numerical example in Fig.~\ref{fig:error-by-layer} for the LFXXZ model with $J_x=J_y=h_z=1.0,J_z=0.3$. The plot shows the absolute error between using $\bar{O}$ as the root observable and tracking the expectation values of other operators that have decreasing weights in the graph and using these lower weight operators directly as the root observable with a correspondingly higher pruning threshold. This shows that the explanation presented here regarding the accuracy of non-root observables in the predefined basis pruning scheme holds. 

\section{1D MFIM with different parameters and observables} \label{app:extra-MFIM}
\begin{figure}
    \centering
    \includegraphics[width=0.5\linewidth]{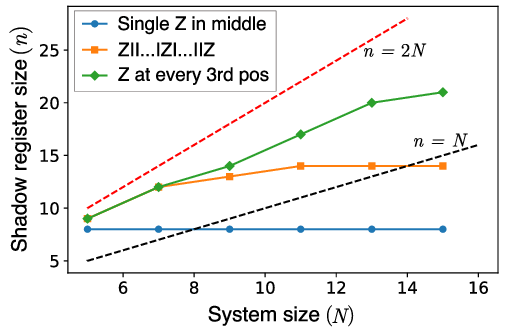}
    \caption{For the mixed field Ising model, the scaling of the shadow register size is shown for increasing system size for different observables at a fixed threshold $\epsilon = h_z^2$ and $h_z = 0.3$. The crossover point to the efficient regime is dependent on the locality and the Hamming weight of the observables. However, the plateauing behavior is common for all the observables. }
    \label{fig:MFIM_diff_obs}
\end{figure}

In this section, we present additional results for approximate shadow Hamiltonia simulation applied to the MFIM for the predefined and hybrid Pauli basis pruning scheme. While the main text focused on single-weight Pauli strings and magnetization observables, here we demonstrate that pruning remains effective for non-local and high-weight Pauli strings as well. However, as the weight and non-locality of the observable increase, the system size at which shadow simulation becomes advantageous also increases. This behavior is illustrated in Fig.~\ref{fig:MFIM_diff_obs}, which shows the scaling of the shadow register size for three different observables: a single Pauli-$Z$ operator acting on the middle qubit, a non-local three-weight Pauli string with $Z$ operators on the first, middle, and last qubits, and a high-weight Pauli string with $Z$ operators appearing at every third site, resulting in a weight that grows with system size. In all cases, the shadow register size eventually saturates, although the crossover to the regime where shadow simulation becomes advantageous occurs at different system sizes. For the observable with system-size-dependent weight, the crossover is not yet visible in the figure, but the flattening of the curve suggests that it will occur at larger system sizes.

We also investigate how the transverse field strength $h_z$ affects the accuracy of the pruned shadow Hamiltonian simulation. To study the effect of $h_z$, we simulate the time evolution of the pruned shadow Hamiltonian for a system size $N=8$ using a cutoff $\epsilon = h_z^3$. This choice ensures that the pruned shadow set $\mathbf{S}_{\bar{O}}$ remains unchanged for the fixed observable $\bar{O} \equiv Z_4$. As expected, the accuracy of the shadow Hamiltonian simulation deteriorates with increasing $h_z$, since operators excluded from $\mathbf{S}_{\bar{O}}$ retain significant weights at larger field strengths and therefore contribute non-negligibly to the dynamics. The accuracy is quantified using the cosine similarity between the shadow-simulation and exact-evolution time series:
\begin{equation}
\text{Accuracy} =
\left|
\frac{\vec{V}_S \cdot \vec{V}_E}
{||\vec{V}_S|| \, ||\vec{V}_E||}
\right| \;, \label{eq:accuracy-def}
\end{equation}
where $\vec{V}_S$ and $\vec{V}_E$ denote the vectors corresponding to the shadow and exact time-evolution data, respectively.
\begin{figure}
    \centering
    \includegraphics[width=0.5\linewidth]{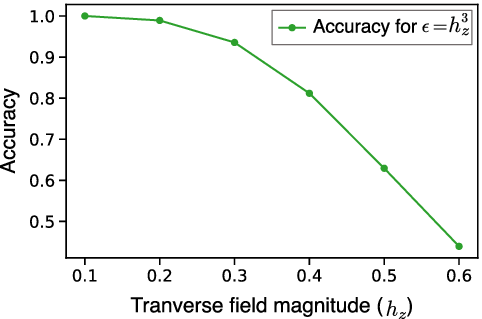}
    \caption{Effect of increasing $h_z$ on the  accuracy of a time evolution simulation for the MFIM with $N=8$ is demonstrated in this figure. The threshold is chosen to be $\epsilon = h_z^3$ such that for each $h_z$, the pruned shadow set remains the same. Time evolution is performed for $t=6$ with time steps $\Delta t = 0.1$ and accuracy is defined in Eq.~(\ref{eq:accuracy-def}).}
    \label{fig:MFIM_diff_xi}
\end{figure}

\begin{figure}
    \centering
    \includegraphics[width=0.5\linewidth]{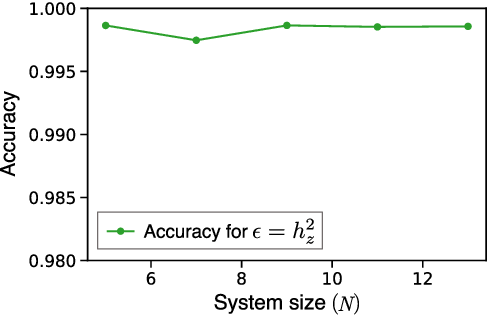}
    \caption{The accuracy of time evolution in the MFIM with increasing system size for a fixed threshold after shadow register size reaches saturation. The time evolution is performed for a threshold $\epsilon = h_z^2$ and observable $\bar{O} =  Z_{N/2}$, the shadow register size saturates at $N_q = 8$ beyond system size $N = 5$. As the system size is increased to $N=13$ spins, the accuracy of the time evolution remains the same. Time evolution is performed till $t=6$ with step size $\Delta t = 0.1$. }
    \label{fig:accuracy-vs-N-postsaturation}
\end{figure}

In Fig.~\ref{fig:accuracy-vs-N-postsaturation}, we show that after the shadow register size saturates for a particular observable and choice of threshold, increasing the system size does not affect the quality of the time evolution.

\begin{figure}
    \centering
    \includegraphics[width=0.5\linewidth]{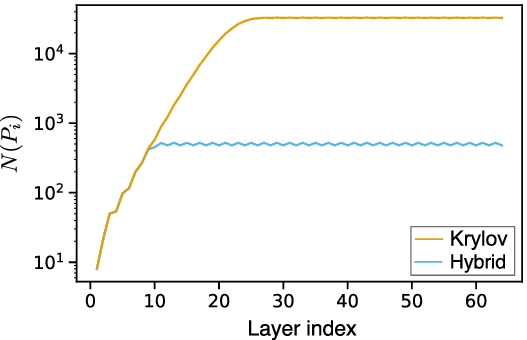}
    \caption{Comparison of number of Pauli strings in each Krylov basis vectors between the Krylov-basis method and the hybrid method for the 1D MFIM with $N=8$ spins and $h_z = 0.3$. For the hybrid method, a predefined basis pruning threshold of $h_z^3$ has been used for the MFIM.}
    \label{fig:pauli_comparison_hybrid_krylov_mfim}
\end{figure}

Finally, to demonstrate the effectiveness of the hybrid method in reducing the memory overhead associated with storing the basis vectors and defining the initial shadow state, we compare the number of Pauli strings appearing in each Krylov basis vector for the pure Krylov method and the hybrid method. This comparison is shown in Fig.~\ref{fig:pauli_comparison_hybrid_krylov_mfim} for the one-dimensional MFIM with $N=8$. Since the pruned set $\mathbf{S}_{\bar{O}}$ contains far fewer Pauli strings than the full Pauli basis, each Krylov vector generated within the hybrid method contains significantly fewer Pauli strings than its counterpart in the pure Krylov approach.

\section{Power law Ising model}\label{app:PLIM}
\begin{figure}[ht!]
    \centering
    \includegraphics[width=0.5\linewidth]{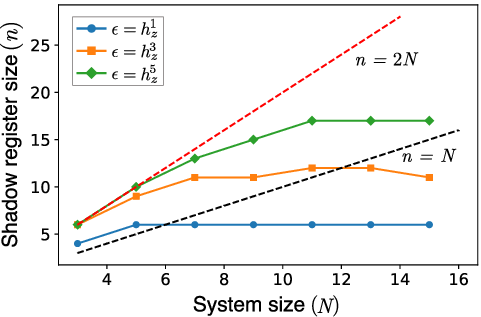}
    \caption{The shadow register size exhibits saturation with increasing system size in the power-law Ising model, similar to the behavior observed in the MFIM. To illustrate this effect, we consider the single-qubit observable $Z_{N/2}$, with Hamiltonian parameters chosen as specified in the main text. The small decrease in the shadow register size for the threshold $\epsilon = h_z^3$ between the 13- and 15-qubit systems arises from the particular random disorder realization, where the disorder amplitudes happen to be smaller in the latter case. The saturation behavior is demonstrated for the thresholds $\epsilon = h_z$, $h_z^3$, and $h_z^5$.}
    \label{fig:PLIM}
\end{figure}

In this section, we show the application of the pruning schemes for a spin-system with all-to-all interaction. We consider the power-law Ising model (PLIM) with disorder defined by the Hamiltonian 
\begin{equation}
    H_{\rm PLIM} = 
J\sum_{i<j}
\frac{X_i X_j}{|i-j|^\alpha}
\, 
\;+\;
h_x \sum_{i=1}^{N} X_i
\;+\;
 \sum_{i=1}^{N} (h_z+D_i) Z_i
\;,
\end{equation}
where $D_i = D\, r_i$, with $r_i \sim \mathrm{Uniform}(-1,1)$. The interaction strength in this model decays as a power law with the distance between interacting spins. In our simulations, we fix the parameters to $\alpha = 2$, $J=1$, and $h_z = D = 0.2$, representing weak transverse and disorder fields, respectively. Depending on the parameter regime, the disordered PLIM exhibits both thermalizing and many-body localized dynamics, making it a versatile and physically rich toy model for studying quantum many-body phenomena \cite{HaukePRB2015}.

Here, our primary interest is to investigate the efficiency of the pruning schemes introduced in this work for spin chains with interactions beyond nearest neighbors. As a non-integrable model with effectively all-to-all connectivity, the full shadow set $\mathbf{S}_{\bar{O}}$ associated with any root observable $\bar{O}$ spans the complete operator space of size $4^N - 1$. To analyze the effect of pruning, we consider the single-site observable $Z_{N/2}$, same as the MFIM. Despite the long-range connectivity, the PLIM displays a saturation of the shadow register size as the system size increases (see Fig.~\ref{fig:PLIM}), although the crossover to this efficient regime occurs at larger system sizes compared to the MFIM.

\section{Error in magnetization plots}\label{app:error-magnetization}
In this section, we show the errors between the magnetization of the 1D and 2D MFIM for different pruning methods. The plots here are complementary to Fig.~\ref{fig:MFIM_main_text} and Fig.~\ref{fig:2DMFIM-magnetization}. The color schemes are the same as the main text, each representing the same thresholds and methods used.

\begin{figure}[ht!]
    \centering
    \includegraphics[width=1.0\linewidth]{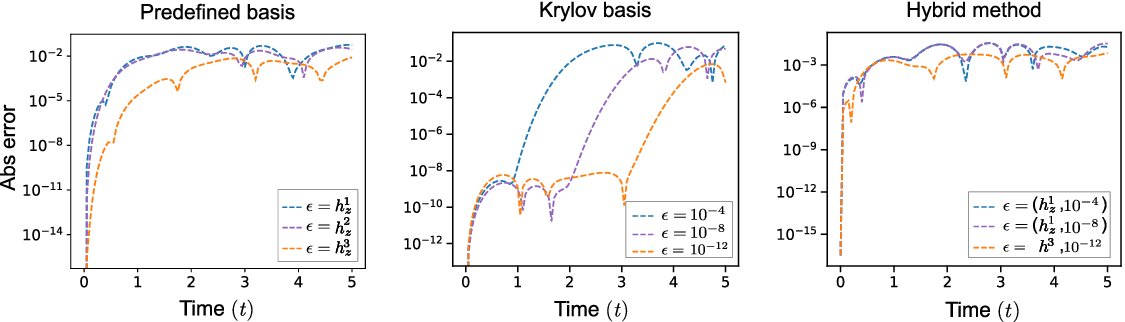}
    \caption{Absolute error in logscale with respect to exact diagonalization of the average magnetization dynamics for 1D MFIM for the same parameters used in the main text. All color schemes and thresholds chosen correspond to the same thresholds in Fig.~\ref{fig:MFIM_main_text}. The three columns represent the three methods.}
    \label{fig:error_MFIM}
\end{figure}

\begin{figure}[ht!]
    \centering
    \includegraphics[width=0.5\linewidth]{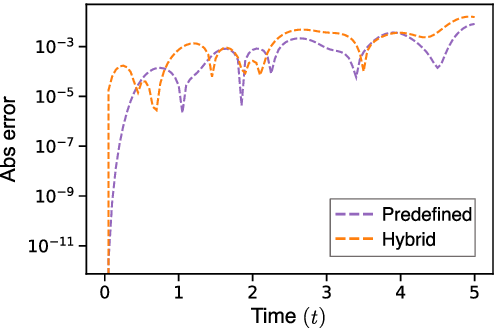}
    \caption{Absolute error in logscale with respect to exact diagonalization of the average magnetization dynamics for the 2D MFIM for the same parameters used in the main text. All colors and pruning schemes chosen correspond to the same ones in Fig.~\ref{fig:2DMFIM-magnetization}.}
    \label{fig:2DMFIM_error}
\end{figure}

\section{Inefficient pruning cases}\label{app:negative-example}
\begin{figure*}[t!]
    \centering
    \includegraphics[width=1.0\linewidth]{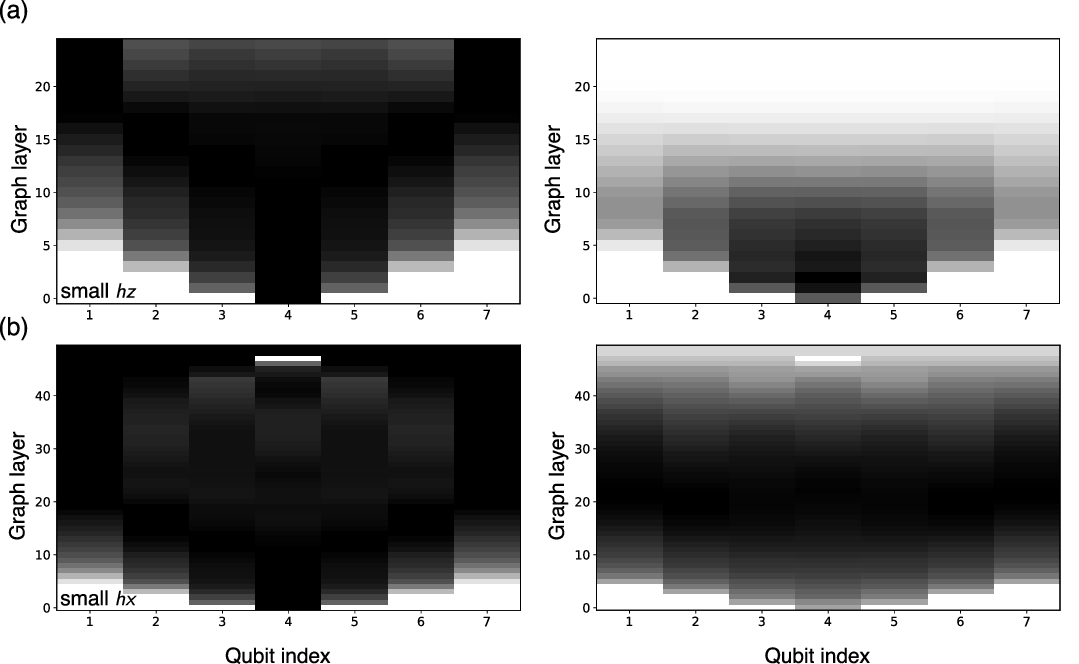}
    \caption{Operator support growth for the predefined-basis pruning scheme in the MFIM with $N=7$ and initial observable $IIIZIII$ is shown for two parameter regimes: (a) small transverse field $h_z$ and (b) small longitudinal field $h_x$. In the left panels, the grayscale at site index $i$ denotes the number of operators with non-identity Pauli support at that site. In the right panels, the grayscale instead represents the total path weight associated with operators that have non-identity support at site $i$. For small $h_z$, the total weight in the upper layers remains negligible at each site, even though many operators populate these layers, as seen in the left panel of (a). This provides a visual criterion for when pruning is expected to be efficient. By contrast, for small $h_x$, the right panel in (b) shows that later layers carry weights comparable to the initial layers, indicating that pruning is inefficient in this regime.}
    \label{fig:operator-growth_doesnt-work-example}
\end{figure*}

The efficiency offered by the pruning schemes discussed in this article depends on the choice of the Hamiltonian and the observable. In the main text we have shown examples where pruning provides significant resource savings. Here, we provide examples of a few cases where pruning is not advantageous. 

For the MFIM Hamiltonian considered in the main text, we choose the perturbative parameter to be associated with the transverse field. This choice enables an efficient truncation of the set $\mathbf{S}$ when the observable of interest is the magnetization. By contrast, if the longitudinal field is chosen to be small compared with the transverse field and the interaction term, pruning becomes inefficient.

This behavior can be understood from the mechanism by which the support of the observable grows. The operator front is advanced by repeated applications of the transverse-field term. For example, an operator initially acting on the $m$-th qubit can spread through the system via paths of the form
\begin{equation}
Z_m \xrightarrow{X_iX_{i+1}} Y_m X_{m+1} \xrightarrow{Z_i} Y_mY_{m+1}
\xrightarrow{X_iX_{i+1}} Y_mZ_{m+1}X_{m+2} \xrightarrow{Z_i} \cdots
\xrightarrow{X_iX_{i+1}} \cdots .
\end{equation}
In the absence of the $h_z$ term, the operator cannot spread beyond the neighboring sites $m-1$ and $m+1$. This is because the terms generated on these sites are Pauli-$X$ operators, which commute with both the longitudinal-field and the interaction terms. Thus, extensive operator growth is enabled by repeated applications of the transverse field.

When $h_z$ is small, each application of this term that advances the operator front suppresses the weight of all subsequently generated terms by an additional power of $h_z$. As a result, operators appearing at later stages of the growth process carry increasingly small weights, making pruning effective. However, when $h_z=\mathcal{O}(J)$, advancing the operator front no longer produces such a hierarchy of weights. The contributions from different layers then remain comparable, substantially reducing the efficiency of pruning.

This mechanism is illustrated in Fig.~\ref{fig:operator-growth_doesnt-work-example}, where we show the growth of an operator initially consisting of a single $Z$ acting on the central qubit. The color density indicates the number of operators in each layer that contain a non-identity Pauli operator on a given qubit. After incorporating the corresponding path weights, we observe that although many operators appear in later layers, their weights are strongly suppressed when their generation requires repeated applications of the small transverse-field term. In contrast, when $h_z$ is of $\mathcal{O}(J)$, the path weights are comparable to the energy scale of the Hamiltonian and different layers of the operator-growth graph contribute with comparable weight (see right panel of Fig.~\ref{fig:operator-growth_doesnt-work-example}(b)).

Another situation in the MFIM where pruning becomes inefficient arises when the observable is a Pauli string whose Hamming weight is comparable to the system size. For instance, for an observable of the form $Z^{\otimes N}$, where $N$ denotes the system size, the dimension of the pruned set scales as $2^{n}$, with $n > N$. This behavior is again closely connected to the operator-growth mechanism discussed above.

\putbib[biblio]
\end{bibunit}

\end{document}